\documentclass[a4paper,12pt]{article}
\usepackage{tabularx} 
\usepackage{amsmath}  
\usepackage{amssymb}
\usepackage{lscape}
\usepackage{caption}

\usepackage{graphicx} 
\graphicspath{{Plots/}}
\usepackage{placeins}
\usepackage[margin=1.09in,letterpaper]{geometry} 
\usepackage{cite} 
\usepackage[final]{hyperref} 
\hypersetup{
	colorlinks=true,       
	linkcolor=blue,        
	citecolor=blue,        
	filecolor=magenta,     
	urlcolor=blue         
}

\newcommand{\mpr}[1]{\textcolor{black}{#1}}

\newcommand{\mppr}[1]{\textcolor{black}{#1}}

\newcommand{\mc}[1]{#1}
\newcommand{\mr}[1]{#1}

\begin{document}

\title{Novel flavour-changing neutral currents\\[0.2cm] in the top quark sector}
\author{Nuno Castro$^a$, Mikael Chala$^b$, Ana Peixoto$^a$, Maria Ramos$^a$\\[0.5cm]
$^a$Laborat\'orio de Instrumenta\c{c}\~ao e F\'isica Experimental de 
Part\'iculas,\\ Departamento de F\'isica, Universidade do Minho, \\
4710-057 Braga, Portugal\\[0.2cm]
$^b$CAFPE and Departamento de F\'isica Te\'orica y del 
Cosmos,\\
Universidad de Granada, E–18071 Granada, Spain
}

\date{}

\maketitle

\begin{abstract}
We demonstrate that flavour-changing neutral currents in the top sector, 
mediated by leptophilic scalars
at the electroweak scale, can easily arise in scenarios of new physics, and in 
particular in composite
Higgs models. We moreover show that such interactions are \mc{poorly} 
constrained by 
current experiments, while
they can be searched for at the LHC in rare top decays and, more generally, in the 
channels $pp\to tS(S)+j$, with $S\to\ell^+\ell^-$. We provide 
dedicated analyses in this respect, obtaining that \mc{cut-off} scales as large 
as $\Lambda\sim$ 90 TeV can be probed with an integrated luminosity of 
$\mathcal{L} = 150$ fb$^{-1}$.
 \\[0.2cm]

Keywords: rare decays, top quark, leptons, flavour-changing neutral currents
\end{abstract}

\newpage


\section{Introduction}
New pseudoscalars $S$ with mass $m_S$ close to the electroweak (EW) scale, singlets of the Standard
Model (SM) gauge group, have two opposing faces.
On one hand, they are predicted in very different and well motivated scenarios of
new physics. These include, among others, the NMSSM~\cite{Ellwanger:2009dp} in which an extra supersinglet
reduces the $\mu$-problem; as well as a big part of the composite Higgs 
models
(CHM) developed to the date~\cite{Contino:2003ve,Gripaios:2009pe,Chala:2012af,Vecchi:2013bja, 
Bellazzini:2014yua,Sanz:2015sua,Ma:2015gra,Belyaev:2015hgo,Chala:2016ykx,
Gripaios:2016mmi,Chala:2018qdf,Chala:2019mve,Cacciapaglia:2019bqz,Ramos:2019qqa}
. Moreover, pseudoscalar singlets have 
been showed to be excellent 
candidates to accommodate EW baryogenesis with two step phase transitions at 
which CP is spontaneously 
broken~\cite{Espinosa:2011eu,Chala:2016ykx,Vaskonen:2016yiu,Beniwal:2017eik,
Kang:2017mkl,Grzadkowski:2018nbc,DeCurtis:2019rxl,Ramos:2019qqa}. Also, they 
have been proved to explain the $g-2$ anomaly of the muon~\cite{Liu:2018xkx}.

However, on the other hand, pseudoscalar singlets around the EW scale are very
difficult to detect. The first reason is that at the renormalisable level they
only interact with
the Higgs boson. If they are above the threshold $\sim m_h/2$, they are 
therefore produced only with extremely low cross section by
means of an off-shell Higgs; being even
out of the reach of a potential 100 TeV collider~\cite{Craig:2014lda,Ruhdorfer:2019utl}. Moreover, the strong
constraints on dipole moments~\cite{Baker:2006ts,Andreev:2018ayy} forbids any 
sizable mixing with the Higgs,
while other production mechanisms are mediated by higher-dimensional operators and
therefore suppressed by the cut-off 
scale~\cite{Franceschini:2016gxv,Gripaios:2016xuo}, that hereafter we 
refer to as $\Lambda$.

In light of these results, there has been research exploring novel
production mechanisms for new pseudoscalar singlets. One of the most exciting possibilities
is producing such particles in the decay of top quarks via effective 
interactions. Such proposal aims to
exploit the huge top quark production rate at the Large Hadron Collider (LHC) and future facilities.
Search strategies for $t\to Sq, S\to b\overline{b} (\gamma\gamma)$ have been
discussed in Ref.~\cite{Banerjee:2018fsx}. (See 
Refs.~\cite{AguilarSaavedra:2000aj,Atwood:2013ica,Papaefstathiou:2017xuv} for 
studies focused on
$m_S\sim m_h\sim 125$ GeV and Refs.~\cite{Aaboud:2017mfd,Aaboud:2018pob} for 
experimental works.) For
$m_S\gtrsim m_t\sim 172$ GeV, the top quark decays non resonantly; being adequately
described by four-fermion interactions~\cite{Chala:2018agk}.

In this article, we extend previous works in this topic in three ways. First, 
we consider
the rare top decay $t\to Sq, S\to \ell^+\ell^-$ including both light leptons and 
taus.
Second, we include the effect of the flavour-violating vertices not only in the decay, but
also in the production of top quarks~\footnote{The interference effects between the production and decay modes were shown to be negligible in Ref.~\cite{Barros:2019wxe}.}. And third, we demonstrate that a more natural
new top decay is $t\to SS, S\to \ell^+\ell^-$ and we also study in detail the LHC reach
to this process.

The text is organized as follows. In section~\ref{sec:framework} we introduce
the relevant dimension-six effective field theory Lagrangian of the SM extended 
with $S$ and compare to 
concrete models. We show which interactions are already constrained by current 
Higgs and flavour data, and define several benchmark points (BP) for 
the subsequent study of top decays. We dedicate sections~\ref{sec:tSll} and 
\ref{sec:tStt} to explore the collider phenomenology of $t\to Sq$ with 
$S\to\ell^+\ell^-$ and $S\to\tau^+\tau^-$, respectively. In 
section~\ref{sec:tSSll} we concentrate on $t\to SSq, S\to\ell^+\ell^-$. We 
conclude in section~\ref{sec:conclusions}.

\section{Interactions and constraints}
\label{sec:framework}

The most generic Lagrangian describing the interactions (that can be induced at tree level in UV completions of 
the SM to dimension six) between a scalar singlet $S$ with mass 
$m_S$ and the SM fields reads~\cite{Franceschini:2016gxv}:
\begin{align}\nonumber
 \Delta L = &- \frac{1}{2}\lambda_{HS} S^2 \left(|H|^2 - \frac{v^2}{2}\right) + 
c_{HS}\frac{(\partial S)^2}{\Lambda^2} |H|^2\\ 
 & + \frac{S}{\Lambda} \mathbf{\overline{f_L}} \mathbf{Y^f} H\mathbf{f_R}  + 
\frac{S^2}{\Lambda^2}\bigg[c'_{HS} |D H|^2  + \tilde{c}_{HS} 
\left(|H|^4-\frac{v^4}{4}\right) + \mathbf{\overline{f_L}} \mathbf{\tilde{Y}^f} 
H\mathbf{f_R}\bigg]\,.
\end{align}
(The addition of the hermitian conjugate in the fermions, as well as 
$\tilde{H}=i\sigma_2 H^*$ when needed, is implied; \mc{with $H$ being the Higgs 
doublet and $\sigma_2$ the 
second Pauli matrix}.) We note that $f$ runs 
over quarks and leptons, $f=q,l$. Let us work in the approximation that the Cabibbo--Kobayashi--Maskawa matrix
is fully diagonal, then no rotation is 
needed on $\mathbf{f_{L,R}}$. The different Wilson coefficients in the 
expression above 
are subject to a number of constraints. Thus, $\lambda_{HS}$, $c_{HS}$ and 
$\tilde{c}_{HS}$ enter the Higgs width: 
\begin{equation}
\Gamma(h\to S S) = \frac{v^2}{32 \pi m_h} \sqrt{1-\frac{4m_S^2}{m_h^2}} 
\left[\lambda_{HS} + c_{HS} \frac{\left(m_h^2 - 2 m_S^2\right)}{\Lambda^2} - 2 
\tilde{c}_{HS} \frac{v^2}{\Lambda^2}\right]^2~.
\label{eq:Hdecay}
\end{equation}
Therefore, values of $m_S < m_h/2$ are \textit{a priori} constrained by LHC 
measurements on the Higgs width, $\Gamma_H \lesssim $ 10 MeV 
\cite{Khachatryan:2016ctc}. This bound 
  can be evaded only if cancellations between the different operators in the 
bracket make this smaller than $\sim 0.05$. Assuming one operator at 
a time, we obtain the bounds $\lambda_{HS} \lesssim 0.05$; 
$c_{HS}/\Lambda^2 \lesssim 3 (5)$ TeV$^{-2}$ for $m_S =  10 (50)$ GeV; and 
${-\tilde{c}_{HS}/\Lambda^2 \lesssim \mr{0.4}}$ TeV$^{-2}$. 

Entries in $\mathbf{Y^f}$ and $\mathbf{\tilde{Y}^f}$ are constrained by  
\textit{e.g.} direct searches for 
resonances~\cite{Sirunyan:2018ikr,Sirunyan:2019vgj}. One important 
exception is entries $i3, 3i$ of 
$\mathbf{Y^q}$. There are no direct limits on these. Moreover, 
indirect constraints from flavour experiments, \textit{e.g.} 
$D^0-\overline{D}^0$ 
oscillations~\cite{Harnik:2012pb,Bona:2007vi,Agashe:2013hma}, involve always 
products of two different Yukawas. They are therefore negligible if 
\mc{\textit{e.g.} the 
entry $13$ or $23$ of $\mathbf{Y^q}$ 
vanishes}~\cite{Banerjee:2018fsx}; same for 
$\mathbf{\tilde{Y}^q}$. However, they can be 
observable in new experiments. Indeed, 
after EWSB and for $m_t > m_S\, (2m_S)$, they lead to signatures such as $t\to 
q S(S)$ arising from
\begin{equation}
 \Delta L \supset \frac{v S}{\sqrt{2}\Lambda} 
\bigg[\mathbf{Y^q}_{i3}\overline{u_L^i} t_R + 
\mathbf{Y^q}_{3i} \overline{t_L} u_R^i 
+\frac{S}{\Lambda}\left(\mathbf{\tilde{Y}^q}_{i3}\overline{u_L^i} t_R + 
\mathbf{\tilde{Y}^q}_{3i} \overline{t_L} u_R^i\right) + \text{h.c.} \bigg]\,,
 \label{eq:lagafterEWSB}
\end{equation}
with $i=1,2$.

\begin{figure}[t]
 \includegraphics[width=0.5\columnwidth]{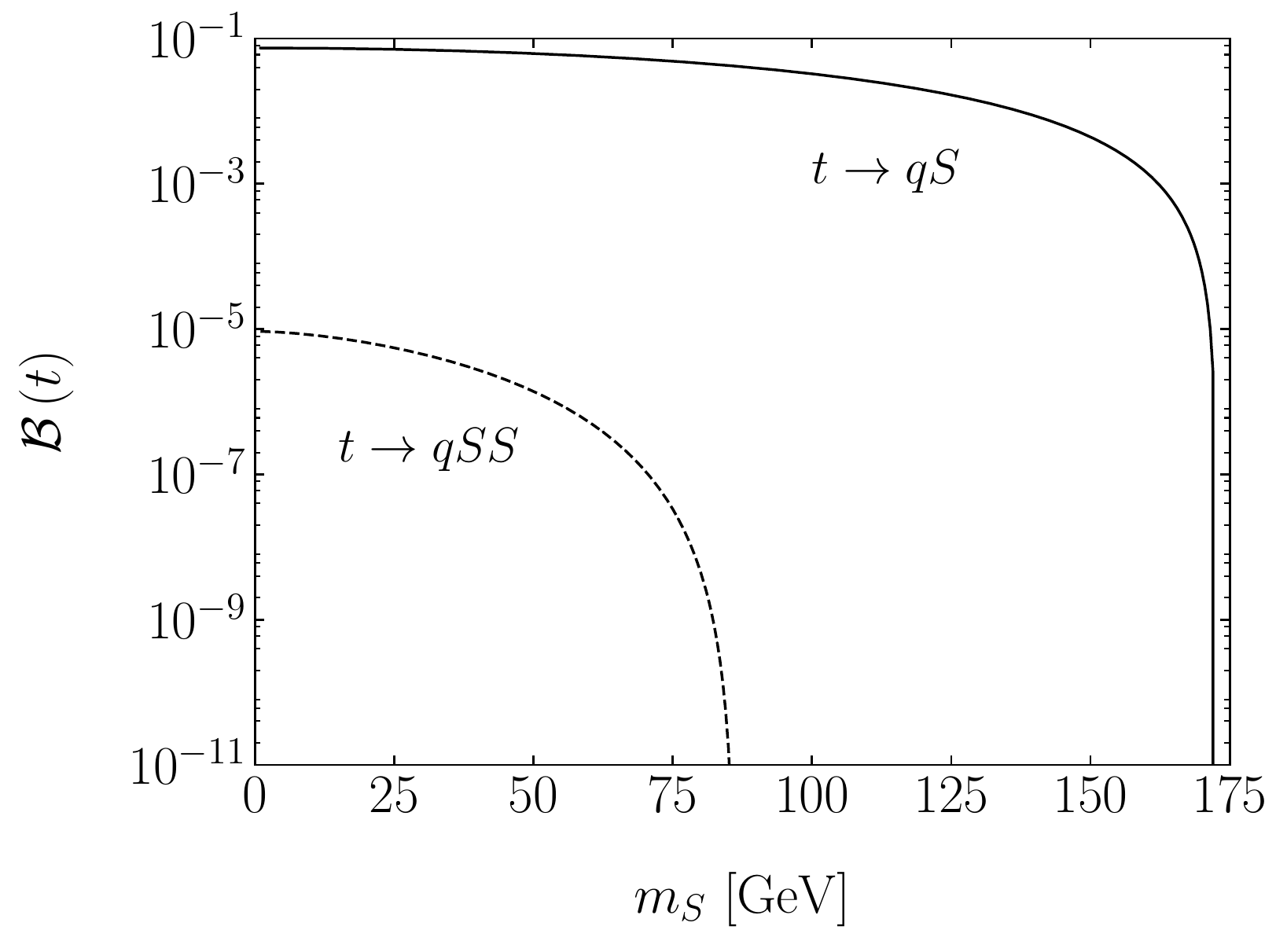}
 \includegraphics[width=0.5\columnwidth]{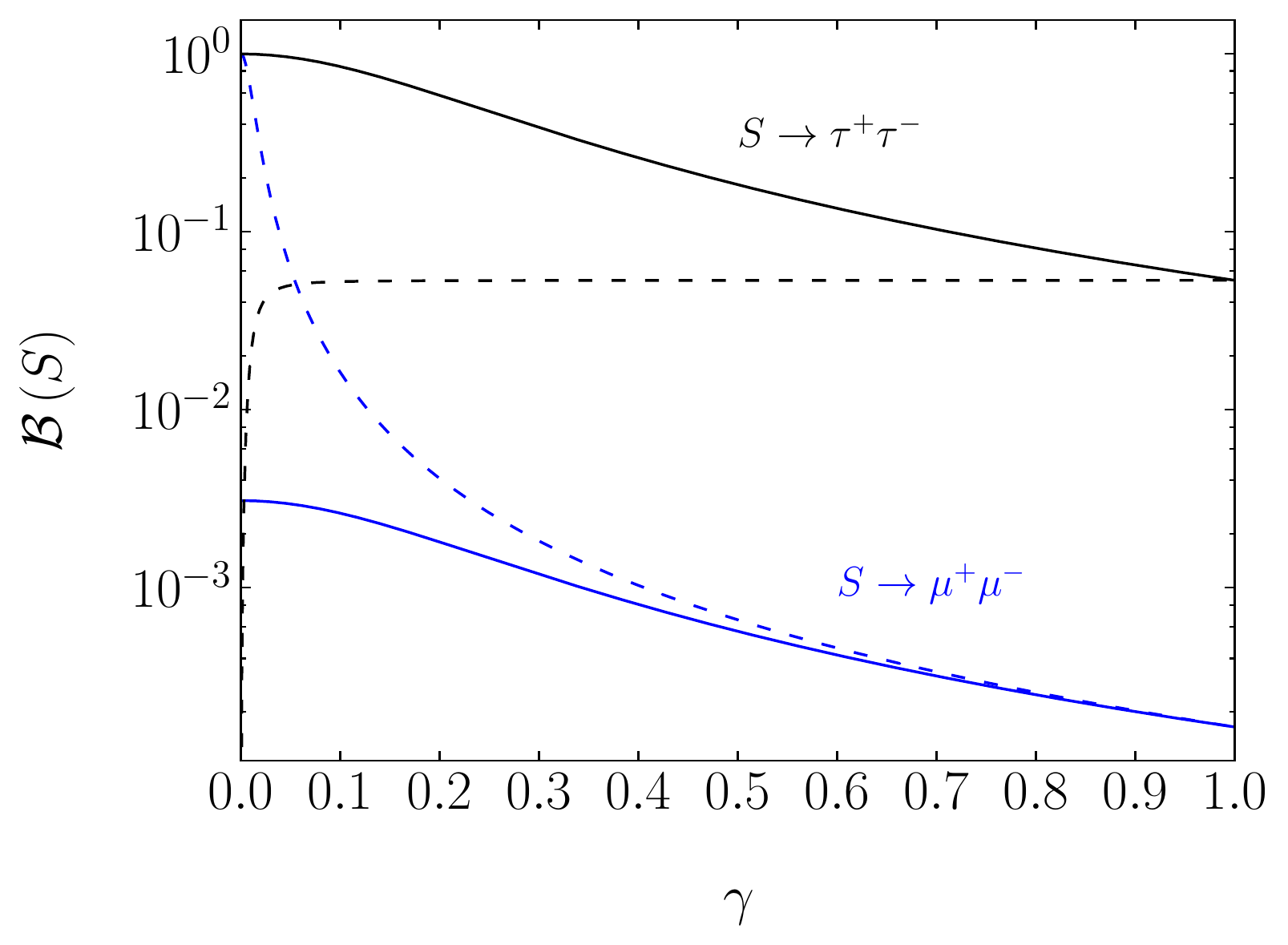}
 \caption{\it Left) Top branching ratios as a function of the mass of $S$, 
for 
$\mathbf{\widetilde{Y}^q}_{i3}=\mathbf{\widetilde{Y}^q}_{3i}=\mathbf{Y^q}_
{i3} = \mathbf{Y^q}_{3i} = 1$ and $\Lambda = 1$ TeV. Right) 
Scalar branching ratios into muons (blue) and taus (black), as a function of the 
$\mathbb{Z}_2$ breaking parameter, $\gamma$, for $m_S = 100$ GeV. We 
represent two cases: (in solid lines) \mr{$\gamma_\ell = \gamma_\tau = 1$} while $\gamma_q = 
\gamma$; and (in dashed lines) \mr{$\gamma_\ell = 1$ and} $\gamma_\tau = 
\gamma_q = \gamma$.
 }\label{fig:decays}
\end{figure}

The decay widths read respectively:
\begin{align}
 \Gamma(t\to q^i S) &= \frac{v^2}{64\pi\Lambda^2} \left[(\mathbf{Y^q}_{i3})^2 + 
(\mathbf{Y^q}_{3i})^2\right] m_t \left(1-x^2\right)^2\,, \\
\nonumber \\
 \Gamma(t\to q^i SS) = \frac{v^2 }{512 \pi^3 \Lambda^4}& 
\left[(\mathbf{\tilde{Y}^q}_{i3})^2 + (\mathbf{\tilde{Y}^q}_{3i})^2\right]  
m_t^3 \bigg[ \frac{1}{3} \sqrt{1 - 4 x^2} \left(1 + 5x^2 - 6 x^4\right) 
\nonumber \\
 & + 2 \left(x^2 - 2 x^4 + 2 x^6 \right) \log{\frac{2 x^2}{1 - 2 x^2 + \sqrt{1 - 4 x^2}}} \bigg]\,,
\end{align}
where we have defined $x = m_S / m_t$. Taking $\Gamma_t \sim 1.4$ GeV as 
reference value of the top width~\cite{Tanabashi:2018oca}, we show in the left 
panel of 
Fig.~\ref{fig:decays} the branching ratio of the top quark into $Sq$ and $SSq$ 
for $\mathbf{\widetilde{Y}^q}_{i3}=\mathbf{\widetilde{Y}^q}_{3i}=$ 
$\mathbf{Y^q}_{i3} = \mathbf{Y^q}_{3i} = 1$ and $\Lambda = 1$ TeV for 
different values of $m_S$. 

The scalar $S$ can subsequently decay into fermions.
In this article we focus on the channel $S\to\ell^+\ell^-$. Assuming that this 
decay mode 
dominates the $S$ width while $t\to q^i S$ requires that only 
$\mathbf{Y^q}_{i3}$ (and/or $\mathbf{Y^q}_{3i}$) and $\mathbf{Y^l}_{jj}$, 
$j=1,2,3$, are non vanishing. This scenario does not easily arise in UV models, 
where diagonal couplings of $S$ to quarks are generally also present, 
proportional to masses, and $b\overline{b}$ dominates the $S$ width; also due to 
the larger number of colours with respect to leptons~\cite{Banerjee:2018fsx}. 
Still, 
the branching ratio to taus is only an order of magnitude smaller. Thus, we 
will consider this unexplored $S$ decay in the context of top flavor-changing neutral currents (FCNCs) in this 
paper. For its cleanness, we will also consider the dimuon channel.

Prospects are very different if $t\to q^i SS$ instead. From the theory point of 
view, it can well be that a $\mathbb{Z}_2$ symmetry $S\to -S$ is only (or 
mostly) broken in the 
lepton sector. Or even just in the muon and electron side; in which case the 
dilepton decay of $S$ is dominant. Let us write $\mathbf{Y^f}_{jj} = 
\gamma_{f_j} y_{f_j}$, where $y_{f_j}$ is the fermion Yukawa and 
$0<\gamma_{f_j}<1$ parameterizes the degree of breaking of the $\mathbb{Z}_2$. 
In the right panel of Fig.~\ref{fig:decays} we show the branching ratio of $S$ 
into taus and muons for different assumptions on this parameter.

The interactions above arise very naturally within 
CHMs~\cite{Kaplan:1983fs,Kaplan:1983sm}, where both the Higgs 
and $S$ are pseudo-Goldstone bosons associated with the spontaneous global 
symmetry breaking $\mathcal{G}/\mathcal{H}$ driven in a new strong sector at 
the confinement scale $\Lambda\sim$ TeV. In these models, the global symmetry 
is only approximate; being explicitly broken by the linear mixing between 
the 
elementary SM fermions and composite operators. (Or equivalently, by embedding 
the SM fermions in incomplete multiplets of $\mathcal{G}$.)

As a matter of example, let us consider 
the next-to-minimal CHM based on the coset 
$SO(6)/SO(5)$~\cite{Gripaios:2009pe}. The generators of $SO(6)$ can be split 
into $SO(5)$ generators, $T$, and coset generators, $X$:
\begin{equation}
 T_{ij}^{mn} = -\frac{i}{\sqrt{2}} (\delta_i^m\delta_j^n - \delta_i^n\delta_j^m)\,\quad X_{ij}^{mn} = -\frac{i}{\sqrt{2}}(\delta_i^m\delta_j^6 - \delta_i^6\delta_j^m)\,,
\end{equation}
with $m<n\in[1,5]$. The pNGB matrix reads
\begin{equation}
 \text{U}=\left[\begin{array}{cccc}
              1_{3\times 3} &  &  & \\
              &1-h^2/(\Lambda^2+\Pi) & -hS/(\Lambda^2+\Pi) & h/\Lambda\\
              &-hS/(\Lambda^2+\Pi)  & 1 -S^2/(\Lambda^2+\Pi) & S/\Lambda\\
              &-h/\Lambda & -S/\Lambda & \Pi/\Lambda^2
             \end{array}\right], ~~ \Pi = 
\Lambda^2\left(1-\frac{h^2}{\Lambda^2}-\frac{S^2}{\Lambda^2}\right)^{1/2}.
\end{equation}
Let us consider the regime in which the left-handed quarks are embedded in the 
representation $\mathbf{6}$, while right-handed up quarks do in both the $\mathbf{6}$ and 
the $\mathbf{15}$. Explicitly:
\begin{align}
 Q_L^I = \frac{1}{\sqrt{2}}(i d_L^I, d_L^I, i u_L^I, -u_L^I, 0,0)\,,\qquad U_{R_1}^I = (0,0,0,0,i\gamma_q u_R^I, u_R^I)\,\\
 \text{and}\qquad U_{R_2}^I = i (T^{12}-T^{34}) u_R^I\,,
\end{align}
with $I$ running over the three quark families and $\gamma_q$ being a positive number. 
To zero momentum and two fermions, only two invariants can be built upon the 
spurions above. One arises from the product of the singlets in the 
decompositions $\mathbf{6}_{Q_L,U_{R_1}} = \mathbf{1}+\mathbf{5}$. The second 
one results from the scalar product of the fiveplets in the decompositions 
$\mathbf{6}_{Q_L} = \mathbf{1}+\mathbf{5}$ and $\mathbf{15}_{U_{R_2}} = 
\mathbf{5}+\mathbf{10}$. Mathematically: 
\begin{align}
 L &= \Lambda y_{IJ}^{(1)} (\overline{U^T Q_L^I})_6 (U^T U_{R_1}^J)_6 - 
\mc{\Lambda} y_{IJ}^{(2)} (\overline{U^T Q_L^I})_m (Tr[U^T U_{R_2}^J U 
X^{m6}])\\
 &=\frac{1}{\sqrt{2}}\overline{\mc{u}_L^I} h u_R^J \bigg[y_{IJ}^{(1)}\left(-1+i 
\gamma_q \frac{S}{\Lambda} + \frac{h^2}{2\Lambda^2}+ \frac{S^2}{2\Lambda^2}\right) + 
y_{IJ}^{(2)} + \cdots\bigg]+\text{h.c.}\,,
\end{align}
with the ellipsis representing terms suppressed by further powers of 
$1/\mc{\Lambda}$. 
Comparing the two equations above with Eq.~\ref{eq:lagafterEWSB}, we find that 
$\mathbf{Y^q}_{ij} \mc{\sim \gamma_q y_{ij}^{(1)}}$ and 
$\mathbf{\tilde{Y}^q}_{ij}\mc{\sim{y_{ij}^{(1)}}}$.  Thus, in 
general, these matrices are not aligned with the Yukawa matrix $\mc{\sim 
y^{(1)}-y^{(2)}}$
and therefore introduce FCNCs.

If the leptons are only embedded in six-dimensional representations we obtain, 
upon rotation: 
\begin{equation}
 L = -\frac{y_l^I}{\sqrt{2}}\overline{l_L^I}h e_R^I \bigg[1-\gamma_l\frac{S}{\Lambda}+\cdots\bigg]+\text{h.c.}\,.
\end{equation}
The assumption that leptons mix with only one representation of the composite 
sector implies that FCNCs vanish; $\mathbf{Y^l}$ is automatically diagonal in 
the physical basis. In fact it is proportional to the lepton Yukawa matrix. 
Thus, the $S$ decay to taus is expected to dominate. 
$S\to\mu^+\mu^-$ should not be neglected, though. First, because 
$m_\mu/m_\tau\sim 0.06$ is not dramatically small. And second because if taus 
couple only to the $\mathbf{15}$, or $\gamma_\tau$ is small, 
then the muon channel dominates \mpr{the} $S$ width\footnote{\mpr{We remark that, in both scenarios with $\gamma_\tau \rightarrow 1$ or $\gamma_\tau \rightarrow 0$ and $\gamma_\mu\to 1$, the flight distance of $S$ is about $10^{-9}-10^{-6}$ cm. Therefore, the singlet decays promptly.}}; we 
refer again to Fig.~\ref{fig:decays}.

We also note that, within the class of CHMs we have just described, there are 
also Higgs mediated FCNCs. Searches for these have been performed in 
\textit{e.g.}
Refs.~\cite{Aaboud:2017mfd,Aaboud:2018pob}. However, as it was first pointed out 
in 
Ref.~\cite{Banerjee:2018fsx}, exploring 
$S$ mediated FCNCs is much more promising for several reasons: \textit{(i)} they 
arise at dimension five, and therefore are less suppressed by powers of 
$v/\Lambda$; \textit{(ii)} the mass of $S$ can lie at values where the SM 
background is less prominent; and \textit{(iii)} contrary to what occurs in the 
case of $S$, there is no parameter space in which the Higgs boson can decay 
sizably into the cleanest final states such as $\mu^+\mu^-$.

Currently, only a few experimental searches are (marginally) sensitive to the interactions discussed before. The first one is the ATLAS search for 
$t\to Zq$ of Ref.~\cite{Aaboud:2018nyl}. In the control region dubbed CR1, this 
analysis 
\mpr{requires} three light leptons (either electrons or muons, denoted 
by $\ell$), two of them  of the same flavour and opposite sign (SFOS), 
exactly one $b$-tagged jet and at 
least two more light jets. Most importantly, the two SFOS leptons with invariant 
mass closer to the $Z$ mass $\sim 91.2$ GeV are required to be out of a 15 GeV mass window
around the $Z$ pole. Consequently $t\to S q$ events with $m_S\neq m_Z$ 
are captured in this region.

The possibility of using this control region 
to constrain interactions not necessarily leading to $t\to Zq$ was first pointed 
out in Ref.~\cite{Chala:2018agk}, which also reports the maximum number of signal 
events allowed by the analysis to be $s_\text{max} = 143$. In order to estimate 
the LHC sensitivity to the proposed signal, we rely on 
home-made routines based on \texttt{ROOT v6}~\cite{Brun:1997pa} with 
\texttt{FastJet v3}~\cite{Cacciari:2011ma}. The simulated events were 
generated with \texttt{MadGraph v5}~\cite{Alwall:2014hca} and 
\texttt{Pythia v8}~\cite{Sjostrand:2014zea}. For the efficiency for selecting 
$t\overline{t}$ events 
in the semileptonic channel we obtain $\epsilon\sim 0.2$. The expected number of 
signal events at $\mathcal{L}=36$ fb$^{-1}$ reads therefore
\begin{equation}
 N \sim 2\times\sigma(pp\to t\overline{t})\times \mathcal{B}(t\to \ell^+\nu b)\times \mathcal{B}(t\to Sq, S\to\ell^+\ell^-)\times \mathcal{L}\times\epsilon,
\end{equation}
with $\mathcal{B}(t\to \ell^+\nu b)\sim 0.27$~\cite{Tanabashi:2018oca} and $\sigma(pp\to 
t\overline{t}) = 832 \pm 29$ pb at 13 TeV~\cite{Czakon:2011xx}. This implies that the upper 
limit on $\mathcal{B}(t\to Sq, S\to\ell^+\ell^-)$ is $\sim 143/(3\times 
10^6)\sim 5\times 10^{-5}$.
This bound in turn translates to a bound on  
$(\mathbf{Y^q}_{i3})^2+(\mathbf{Y^q}_{3i})^2 \lesssim 10^{-4} 
\left(\Lambda/v\right)^2$, (for $m_S \sim m_t/2$).
To the best of our knowledge there are no relevant constraints on the tau channel.

Another analysis sensitive to the proposed top interactions is the ATLAS search for SUSY in 
multilepton final states~\cite{Aaboud:2018zeb}. 
 The maximum number of allowed signal events in this analysis 
for $L = 150$ fb$^{-1}$ is larger than $20$. Upon implementing only a few 
of the cuts, we have checked that such big number arises within our 
framework only if FCNC couplings are larger than $1$ for $\Lambda=1$ TeV. It 
will become clear in the following sections that the dedicated analyses we 
propose are sensitive to a much larger region of the parameter space.

Other similar searches suffer from the same problem. Namely, they are too broad 
in scope and therefore the background is large enough to hide 
the signal we are interested in. Consequently, dedicated searches are required and we discuss 
three examples in very detail in the subsequent sections, aiming to explore 
$pp\to tS+j, S\to\mu^+\mu^-$; $pp\to tS+j, S\to\tau^+\tau^-$ and $pp\to tSS\mr{+j}, 
S\to\mu^+\mu^-$. Note hence that the top FCNCs can be either in the decay of 
the top quark when is pair-produced via QCD (with the extra jet from 
radiation), or directly in the core of $tS$ associate production.

In light of the discussion above, we assume hereafter that $\Lambda \gtrsim$ 
TeV and 
all Wilson coefficients vanish with the exception of~\footnote{Note that for 
$10\,\text{GeV}<m_S<100\,\text{GeV}$ current data from ATLAS, CMS and BaBar 
only constrain $\mathbf{Y^l} \gtrsim 0.1$~\cite{Liu:2018xkx} for $\Lambda = 1$~TeV. Even 
for much smaller values, $S$ decays promptly. \mppr{We remark that LEP bounds at the $Z$ pole \cite{Adam:1993xs} are more than one order of magnitude weaker than the previous constraints.}} 
$\mathbf{Y^l}_{22}$ (or $\mathbf{Y}^l_{33}$) and 
either  $\mathbf{Y^q}_{13}$ or 
 $\mathbf{Y^q}_{23}$ or 
$\mathbf{\tilde{Y}^q}_{13}$ (depending on which process we study).  
We quantify the results in terms of seven benchmark masses: $m_S =$ 20, 50, 80, 
90, 100, 120 and 150~GeV. Given the reduced phase space for $t\to SSq$, we include in addition the benchmark masses $m_S = $ 30, 40, 60 and 70~GeV in the analysis for this channel.
The reach 
of the dedicated analyses proposed below will be compared to the 
following Benchmark Points (BP): 
\begin{align}
&{\rm BP~1}:\quad {\rm \mathbf{Y^q}}_ {i3} = {\rm \mathbf{Y^q}}_ {3i} = 0.01~, \quad \Lambda=~5~{\rm TeV} \quad 
\Longrightarrow \quad \mathcal{B}(t\rightarrow S q) \sim 
10^{-8}-10^{-7}\,,\nonumber\\
&{\rm BP~2}:\quad {\rm \mathbf{Y^q}}_ {i3} = {\rm \mathbf{Y^q}}_ {3i}  = 0.10~, \quad \Lambda =~5~{\rm TeV}\quad 
\Longrightarrow \quad \mathcal{B}(t\rightarrow S q) \sim 
10^{-6}-10^{-5}\,,\nonumber\\
&{\rm BP~3}:\quad {\rm \mathbf{Y^q}}_ {i3} = {\rm \mathbf{Y^q}}_ {3i}  = 0.10~, \quad \Lambda =~1~{\rm TeV} \quad 
\Longrightarrow \quad  \mathcal{B}(t\rightarrow S q) \sim 
10^{-4}-10^{-3}\,,\nonumber\\
&{\rm BP~4}:\quad {\rm \mathbf{ \tilde{Y}^q}}_ {i3} = {\rm \mathbf{ \tilde{Y}^q}}_ {3i}  = 1.00~,\quad \Lambda =~5~{\rm TeV}\quad 
\Longrightarrow \quad \mathcal{B}(t\rightarrow S S q) \sim 10^{-11}-10^{-8}\,,\nonumber\\
&{\rm BP~5}:\quad {\rm \mathbf{ \tilde{Y}^q}}_ {i3} = {\rm \mathbf{ \tilde{Y}^q}}_ {3i}  = 0.20~,\quad \Lambda =~1~{\rm TeV}\quad 
\Longrightarrow \quad \mathcal{B}(t\rightarrow S S q) \sim 10^{-10}-10^{-7}\,,\nonumber\\
&{\rm BP~6}:\quad {\rm \mathbf{ \tilde{Y}^q}}_ {i3} = {\rm \mathbf{ \tilde{Y}^q}}_ {3i}  = 1.00~,\quad \Lambda =~1~{\rm TeV}\quad 
\Longrightarrow \quad \mathcal{B}(t\rightarrow S S q) \sim 10^{-8}-10^{-5}\,,
\end{align}
with $i=1,2$. The range in the branching ratio ensues from the range of 
values of $m_S$.
\begin{figure}
 \centering
  \includegraphics[width=0.9\textwidth]{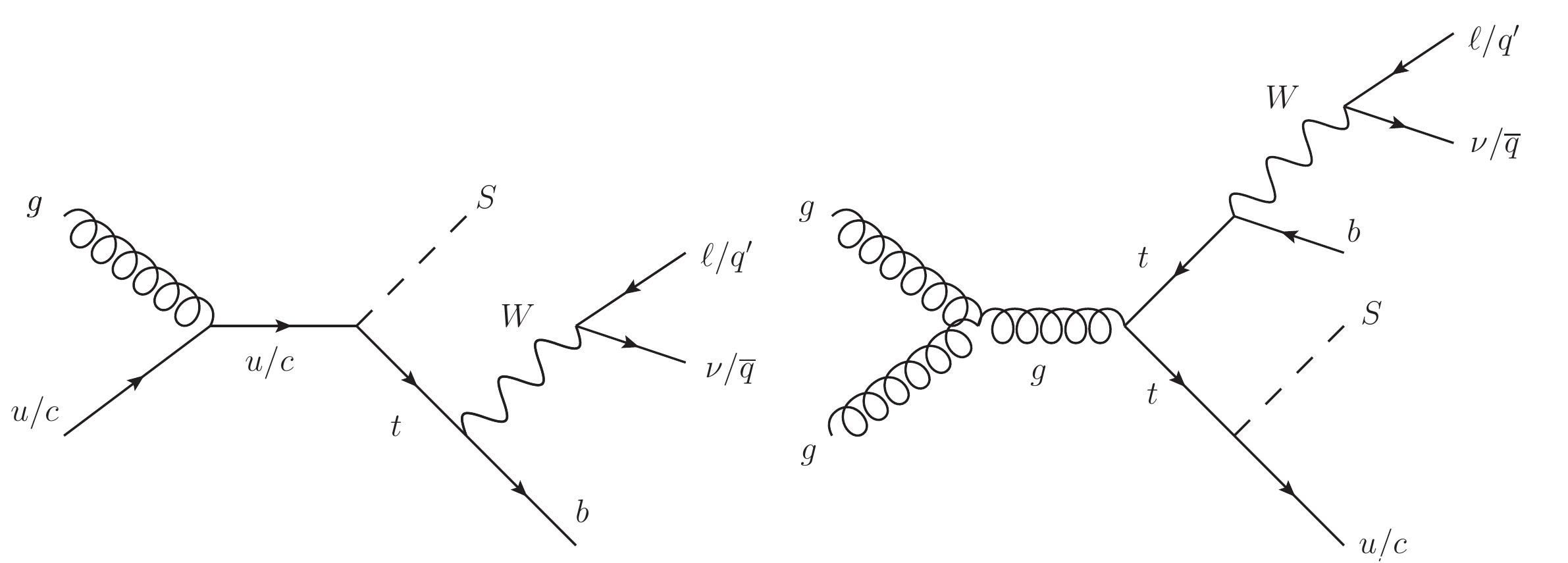}
 \caption{Representative Feynman diagrams for the production of a single top 
quark in association with $S$ via an FCNC interaction (left) 
and for the top-quark pair production with an FCNC top-quark decay into the 
extra singlet (right). \mpr{Similar diagrams but involving two rather than one $S$ in the new physics vertex hold for the production of a top quark in association with $SS$.}}
 \label{fig:diagrams}
\end{figure}

\section{Search for $t\to Sq, S\to\mu^+\mu^-$}
\label{sec:tSll}

The singlet $S$ can arise either in the production or in the decay of the top 
quark; 
see the diagrams on figure~\ref{fig:diagrams}.  This leads to a final state 
with 
exactly one $S$, one top quark decaying into $Wb$ and eventually an additional 
light quark $q = u,c$. 
In this first analysis we study the scenario where $S$ decays into a 
pair of 
muons.
We focus on the leptonic decay of the $W$. Hence, at the 
detector level, we expect three charged leptons, several jets (at least one 
originated by a $b$ meson) and significant missing energy.

We generate signal and background events at $\sqrt{s}=13$ TeV with 
\texttt{MadGraph v5}~\cite{Alwall:2014hca}, with signal model being 
implemented in \texttt{Feynrules v2}~\cite{Alloul_2014}. We subsequently use 
\texttt{Pythia v8} \cite{Sjostrand:2014zea} for simulating the initial and final 
state radiation, the parton shower and the hadronization. At parton level, 
only leptons and photons with a transverse momentum higher than $10$~GeV are 
considered. For the jets, this cut rises to
$20$~GeV. Concerning the absolute pseudo-rapidity $|\eta|$, jets can have a 
value of this variable lower than $5$ while for leptons and photons it should 
be lower than $2.5$. 

We use the Parton Distribution Functions \texttt{NNPDF23LO}~\cite{PDF} and set the 
renormalization and factorization scales to  the default dynamical 
\texttt{MadGraph} value. The total background comprises samples 
from $tW$, $t\Bar{t} V$, $VV$, $ZVV$, $t\Bar{t}$, $V$ + jets and $tZ$, with $V 
= W, Z$. 
For the present analysis, the most dominant background comes from $tZ$ and 
$t\overline{t}$ production, where $t\rightarrow Wb$ and all gauge bosons are 
assumed to decay into muons. Such exclusive samples are motivated by the 
targeted trilepton final state of the signal. (Although the $tW$, diboson and $Z$ + jets 
processes have the largest cross sections, 
they become irrelevant after the cuts on the number of reconstructed leptons and 
jets; check Tab.~\ref{tab:First_Background}.) 

We use \texttt{Delphes} \cite{de_Favereau_2014} to simulate the detector effects 
with the default CMS detector card.
An electron (muon) is considered to 
be isolated if the sum of the transverse momenta of all particles above 
$p_T^{min} = 0.5$ GeV that lie within a cone of radius $R = 0.5$, normalized to 
the lepton $p_T$, is smaller than 0.12 (0.25).

Jets are defined using the anti-$k_t$ algorithm~\cite{Cacciari:2008gp} 
with a radius parameter of 
$R =0.5$. All the jets are required to have $p_T>25$ GeV and to lie within a 
pseudorapidity range of $|\eta|<2.5$.  Leptons must have $p_T > 15$ GeV and 
$|\eta| < 2.4~(2.5)$ for muons (electrons); the hardest lepton is also required 
to have $p_T > 25$ GeV. The effect of the previous cuts on the transverse momentum and pseudorapidity, together with the requirement of exactly three isolated leptons, can be found in the yields tables labeled as ``basic''.

We then select events with at least one 
jet, one of 
them required to be 
tagged as a $b$-quark. The \texttt{Delphes} CMS card was used to parameterize the $p_T$ 
dependent tagging efficiencies for jets initiated by $b$-quarks, as well as to take into account the mistag probability. As an example, 
for a $b$-jet with a transverse momentum of $30$~GeV, the tagging efficiency is 
$55\%$ 
and the mistag rate for a $c$-jet with the same $p_T$ is $12\%$. The scalar 
resonance, $S$, is reconstructed from the hardest $\mu^+\mu^-$ pair (if there is 
none, the event is discarded).
\begin{figure}[t]
 \centering
  \includegraphics[width=0.49\textwidth]{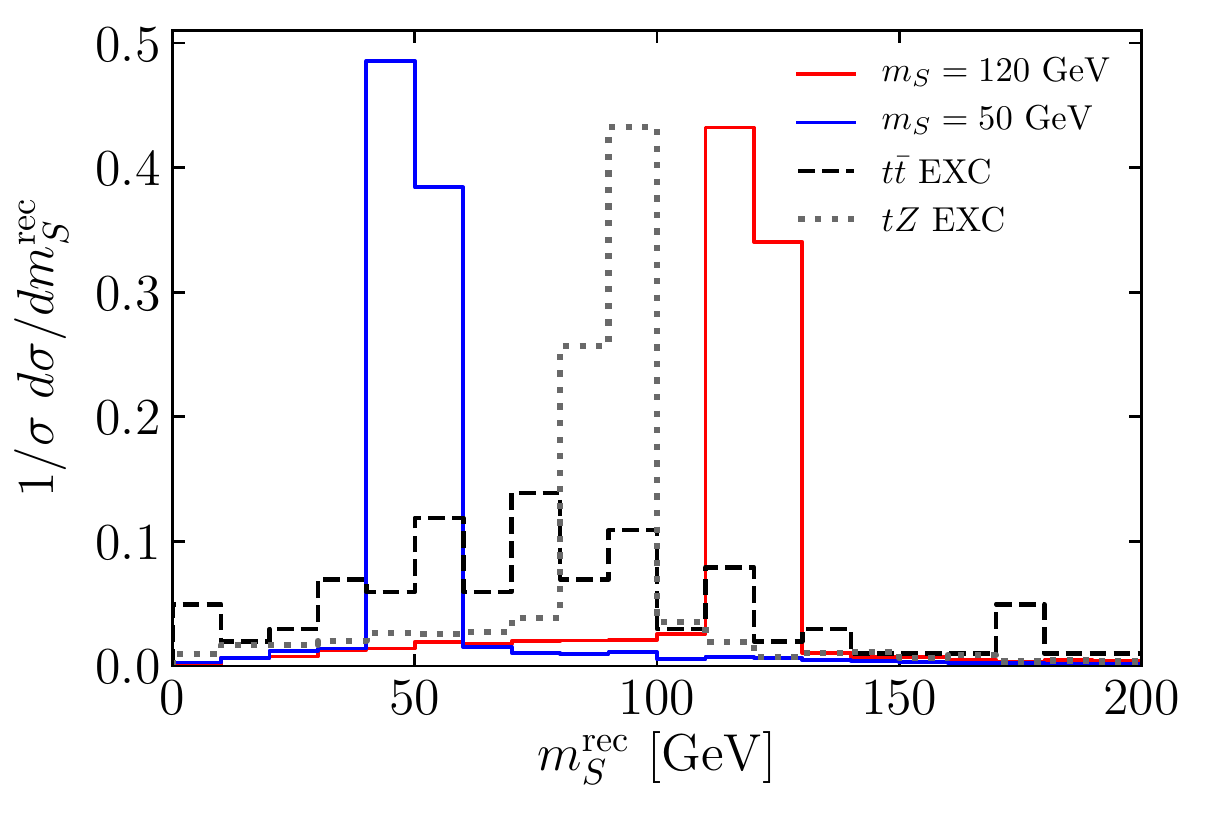}
  \includegraphics[width=0.49\textwidth]{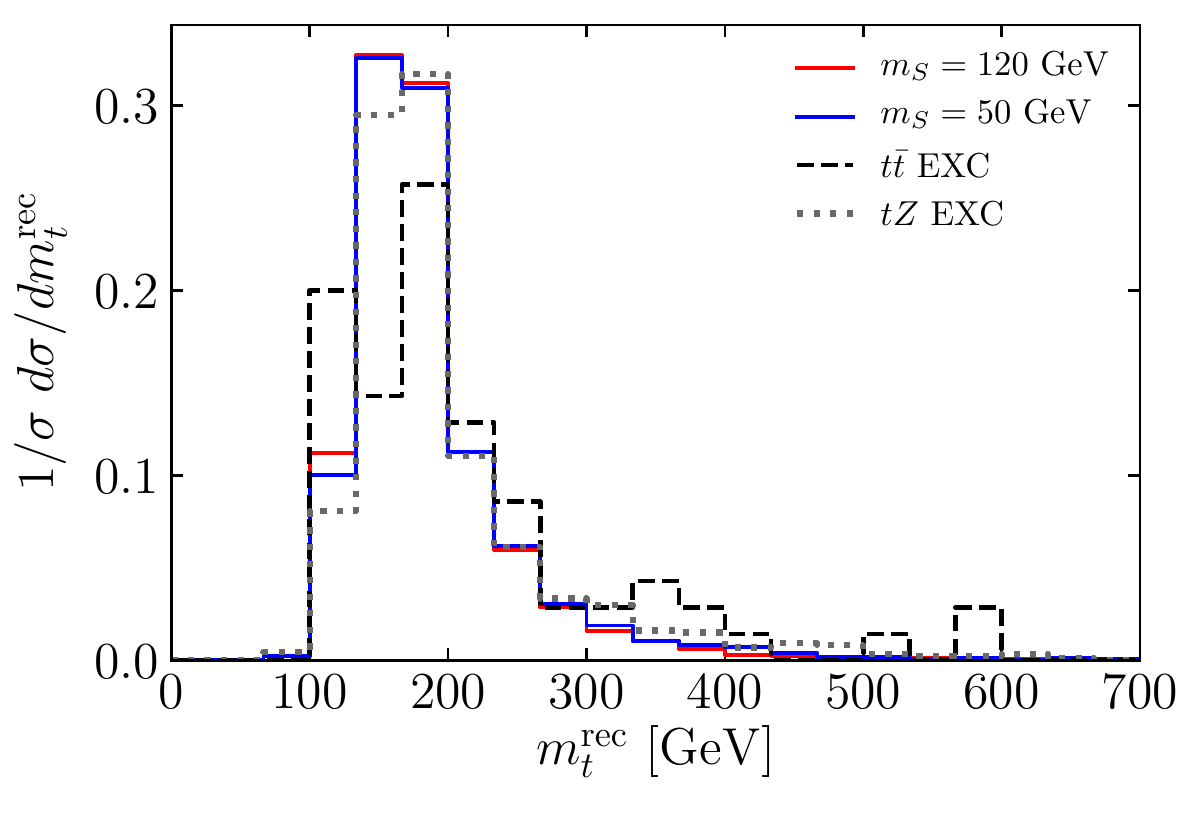}\\
 \includegraphics[width=0.49\textwidth]{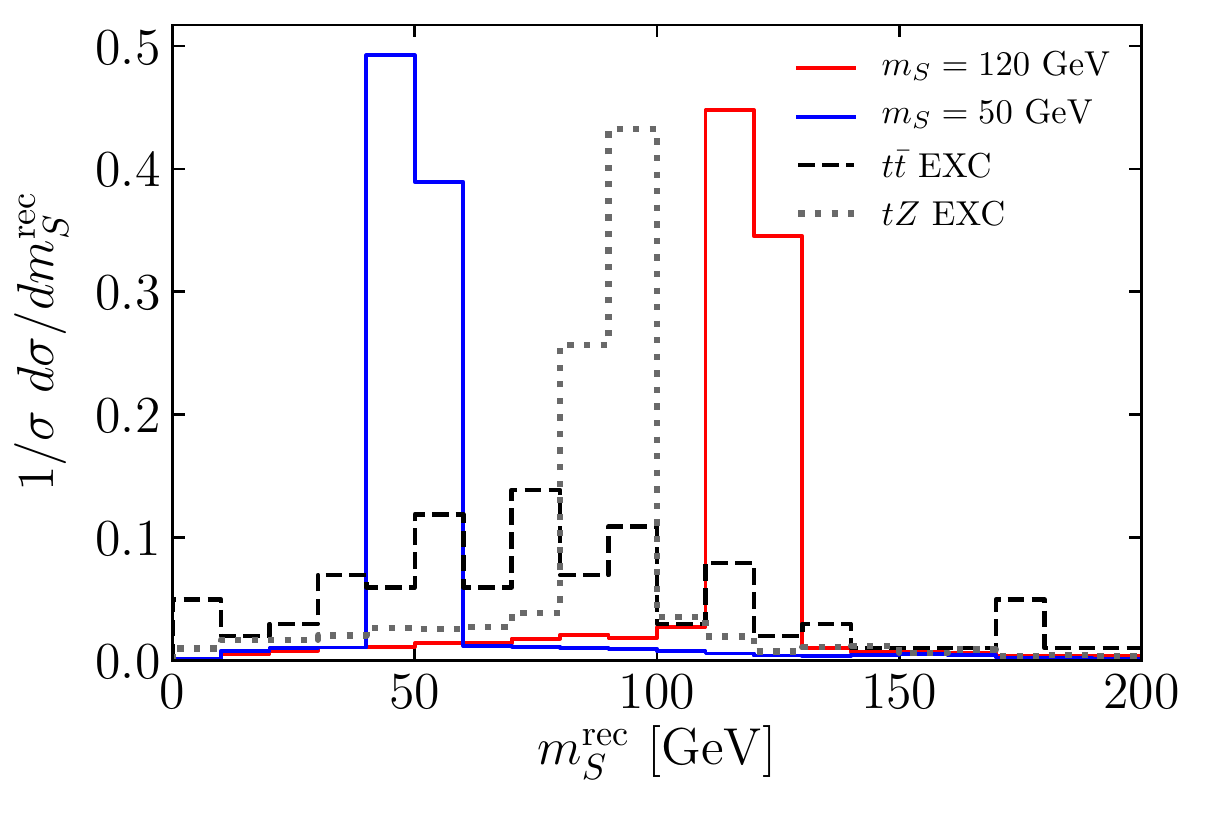}
  \includegraphics[width=0.49\textwidth]{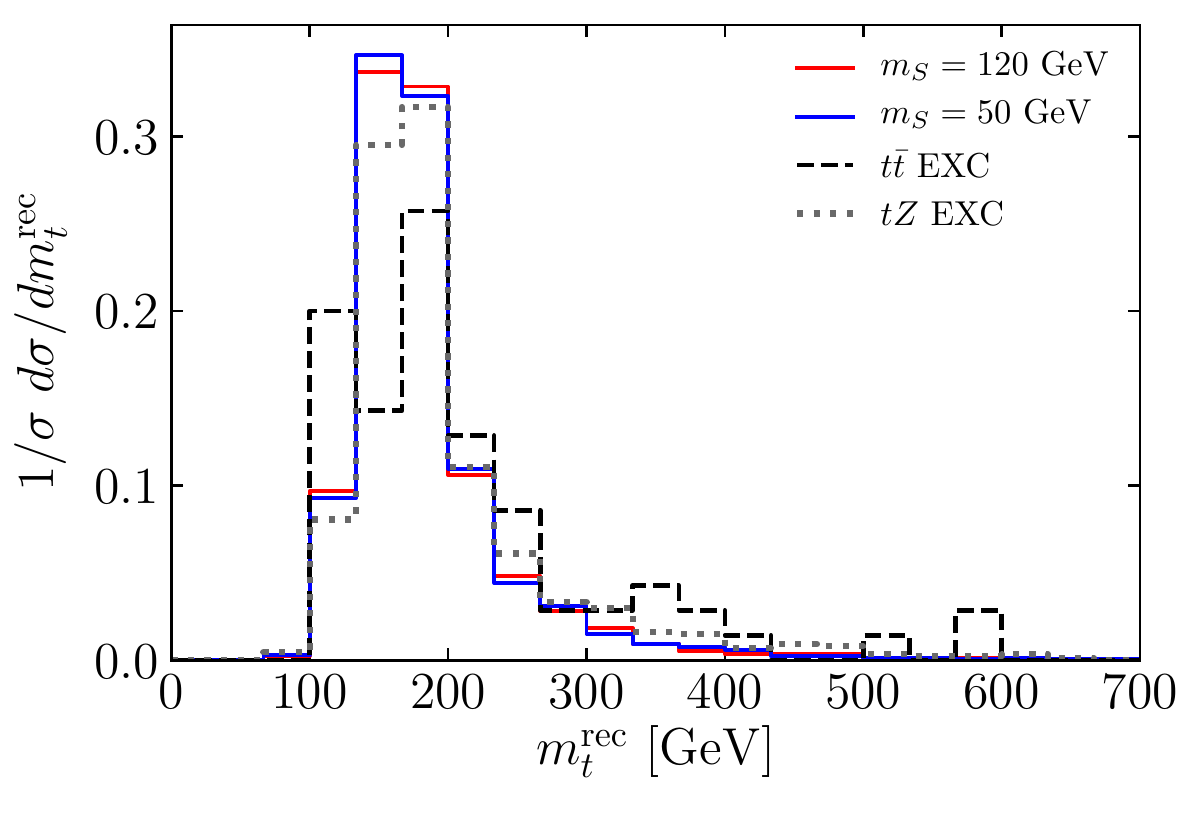}
 \caption{The reconstructed scalar (left) and top (right) mass in the analysis 
proposed for $t\to Sq$, $S\to\mu^+\mu^-$. In the upper (bottom) panel, $q = 
c~(u)$. We represent the distributions of two signal benchmark 
points and the two major background components, after the cut on the particle 
multiplicities; the background samples are generated exclusively, $i.e.$ only 
gauge boson decays into muons are included. The distributions assume a 
collected luminosity of $\mathcal{L} = 150~{\rm fb}^{-1}$. 
 }
 \label{fig:First_mrec}
\end{figure}
The longitudinal component of the missing neutrino four-momentum ($p_\nu$) is 
reconstructed by demanding $m_W^2 = (p_\ell + p_\nu)^2$, where $p_\ell$ 
is the four-momentum of the lepton not coming from the scalar decay and $m_W = 81.2$~GeV is our reference value for the $W$ boson mass. Among the two 
possible solutions, we use the one with smaller absolute value.

Both $p_\nu$ 
and $p_\ell$ are then added to the four-momentum of the $b$-jet to reconstruct the 
SM top quark; its invariant mass being dubbed $m_{t}^{\rm rec}$.
We show the distributions of the scalar and top reconstructed masses for two 
signal benchmark points and for the relevant background components in 
Fig.~\ref{fig:First_mrec}. \mc{The label \texttt{EXC} manifests that the 
corresponding backgrounds are generated assuming the exclusive leptonic mode.}
We require the $m_{t, {\rm rec}}$ variable to be within a window of $50$ GeV 
from the reference top mass $m_t = 172.5$ GeV. 
We impose an additional cut of $1$ TeV on the maximum invariant mass of the 
total system, $m_{\rm total}$, in order to stay in the regime of validity of 
the effective field theory. The impact on the expected signal yield caused by this additional 
requirement is minor when compared \mc{to the other selection cuts,} and 
increases for higher masses (varying between 1.5\% and 7\%  for masses of 
the scalar of 20~GeV and 150~GeV, respectively). The final cut requires the scalar $S$ candidate invariant mass, $m_S^{\text{rec}}$, to be within a mass window of $\pm 30$ GeV 
around the probed value of $m_S$.

The cut flow for the signal with an up and with a charm quark is given in 
Tabs.~\ref{tab:First_Signal_Up} and \ref{tab:First_Signal_Charm}, 
respectively. The scalar mass-independent yields for the background 
components are shown in Tab.~\ref{tab:First_Background} while the 
mass-dependent one is given in Tab.~\ref{tab:First_Background_MassCut}. 
An integrated luminosity of $150~{\rm fb}^{-1}$ is considered for this analysis.

Upper limits on the signal cross section are obtained under the signal absence hypothesis, using the CL$_{\rm s}$ method~\cite{Read:722145}. For this, the distribution of the invariant mass of the reconstructed scalar $S$ after all selection cuts is fitted with \texttt{OpTHyLic}~\cite{Busato:2015ola}. A total of $20$ bins per signal point are considered and Poissonian statistical 
uncertainties on each bin of the distributions are included in the 
computation. An expected upper limit on the signal strength, $\sigma_{95\%}/\sigma_{\rm th}$ ($pp \to tS(q)$, $S \to \mu^+\mu^-$), at $95\%$ confidence level (CL) is then obtained. The 
signal cross section, $\sigma_{\rm th}$, is computed with \texttt{MadGraph v5}. The $\pm1\sigma$ 
and $\pm2\sigma$  variations are also computed, taking into account the statistical uncertainty arising from finite Monte Carlo samples.

We present the results in  
Fig.~\ref{fig:BR_mumu}, where we show the 95\% CL upper limits on the top 
branching ratio
{$\mathcal{B}\left(t\to S q, S\to\mu^+\mu^-\right)$ and cross section 
$\sigma(pp \to t S (q), S \to \mu^+\mu^-)$. The $\pm 1 \sigma$ and $\pm 2\sigma$ 
bands are plotted in green and yellow, respectively.

Note that the sensitivity worsens at $m_S \sim m_Z$ due to larger impact of 
the backgrounds with $Z$-bosons. This effect, however, is attenuated by two factors: 
\textit{(i)} the mass window around the singlet mass in which the events are 
selected is relatively large; and \textit{(ii)} while the $tZ$ background 
distribution is larger around $90$ GeV, the combination of the $tZ$ and $t 
\overline{t}$ background components is equally important in the neighboring bins 
of the reconstructed scalar mass distribution.

\begin{figure}[t]
 \centering
  \includegraphics[width=0.49\textwidth]{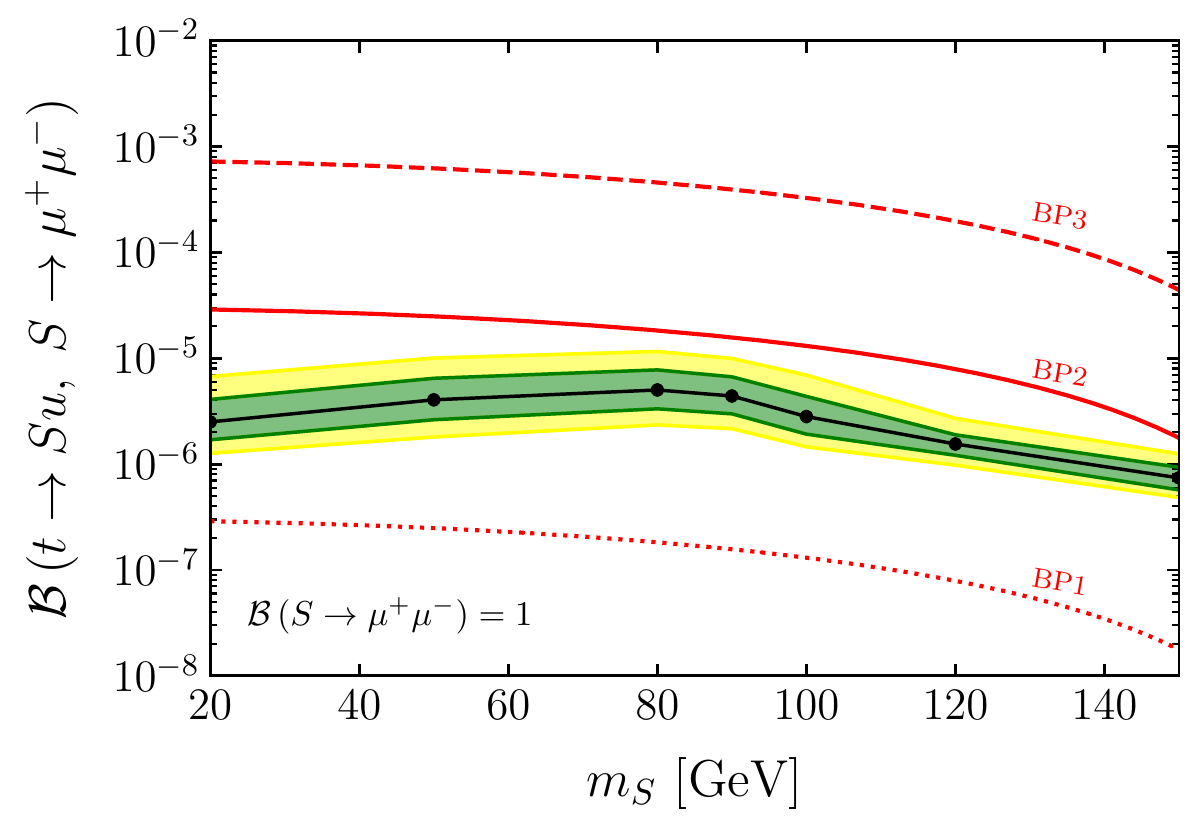}
  \includegraphics[width=0.49\textwidth]{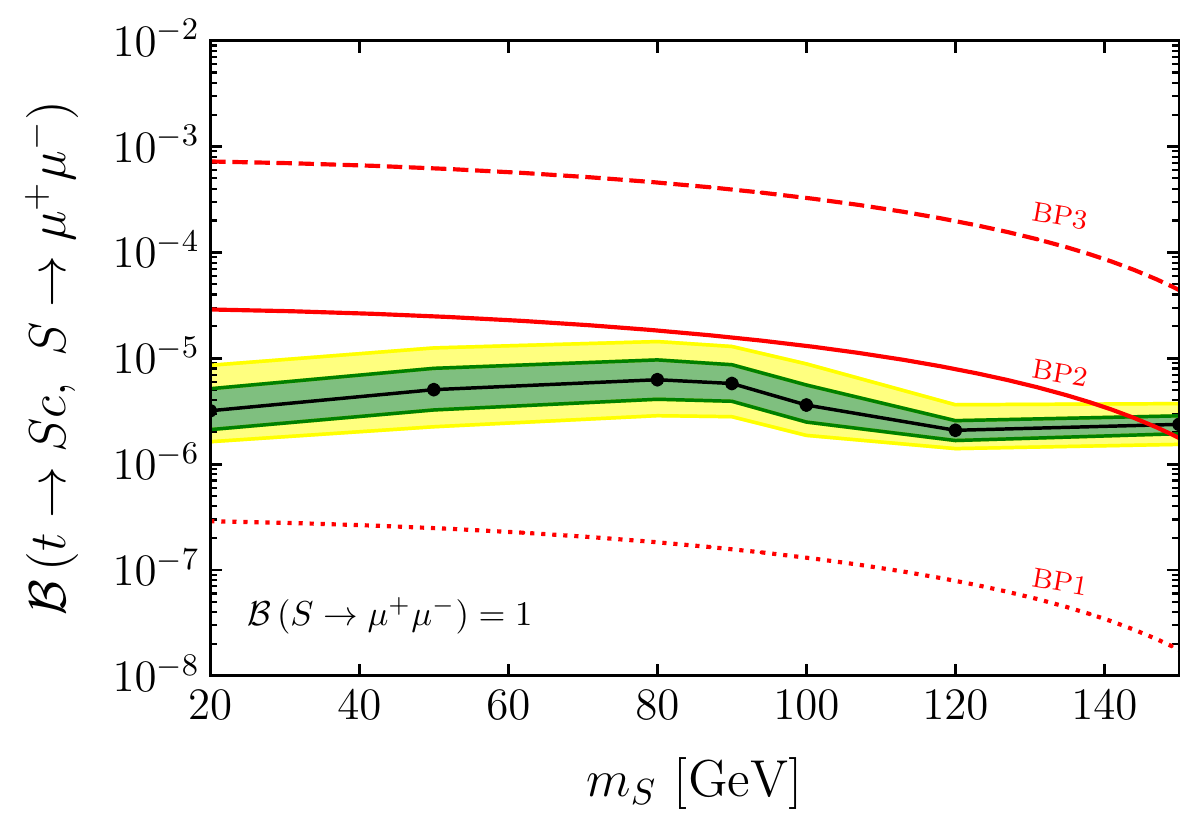}\\
    \includegraphics[width=0.49\textwidth]{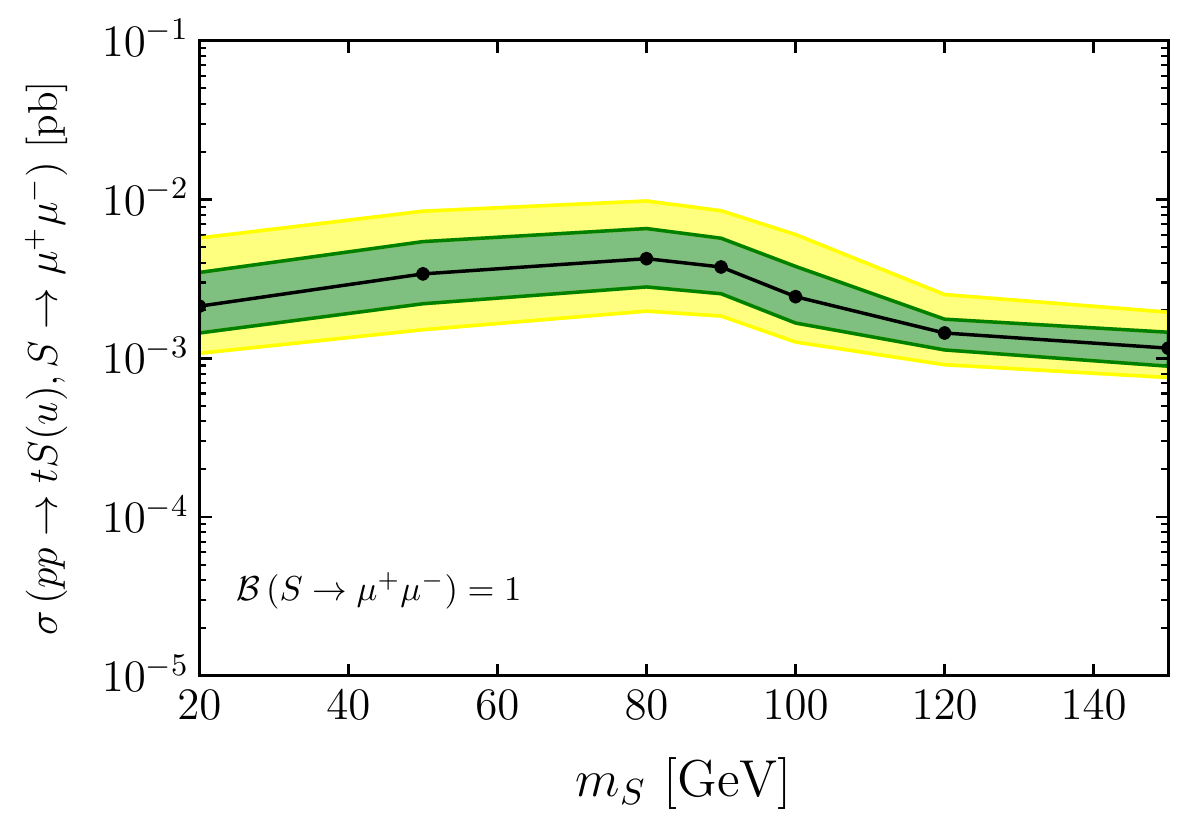}
  \includegraphics[width=0.49\textwidth]{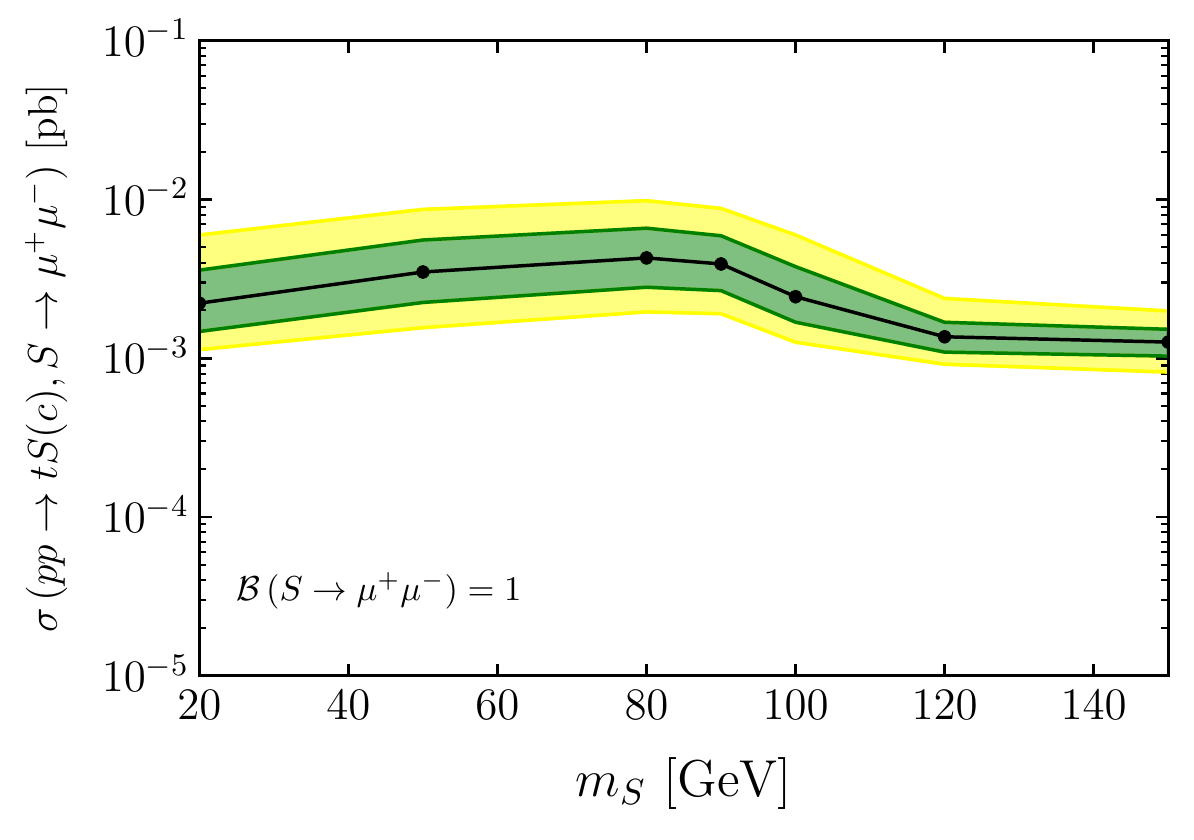}\\
 \caption{In the upper (bottom) panels, we show the 95\% CL limits 
 on the branching ratio (cross section times branching ratio) that can be 
tested in the $\mu^+\mu^-$ channel, in the 
analysis proposed for $t\rightarrow S q,~S\rightarrow \mu^+\mu^-$, with $q = 
u~(c)$ in the panels on the left (right). The limits are obtained for a 
collected luminosity $\mathcal{L} = 150~{\rm fb}^{-1}$. The green and yellow 
bands 
show the  $\pm 1 \sigma$ and $\pm 2\sigma$ uncertainty on the limits, 
respectively. Superimposed are the 
theoretical expectations in three BPs.
}
 \label{fig:BR_mumu}
\end{figure}
\begin{landscape}
\begin{table}[t]
	\centering
	\scalebox{0.75}{
	\begin{tabular}{|l|c|c|c|c|c|c|c|}
		\hline
		Cuts/\mr{$m_S$} & 20 GeV & 50 GeV & 80 GeV & 90 GeV & 100 GeV & 120 GeV & 150 GeV\\\hline
		basic & $6092\pm61$ & $5597\pm54$ & $4769\pm43$ & $4295\pm38$ & $3642\pm32$ & $2416\pm21$ & $899\pm8$ \\\hline
		$n_j > 1$ & $6033\pm60$ & $5537\pm53$ & $4708\pm42$ & $4249\pm37$ & $3589\pm32$ & $2378\pm21$ & $865\pm8$ \\\hline
		$n_b = 1$ & $3249\pm44$ & $2961\pm39$ & $2543\pm31$ & $2301\pm27$ & $1914\pm23$ & $1265\pm15$ & $479\pm6$ \\\hline
		$n_{\mu^+ \mu^-} = 1$ & $3247\pm44$ & $2959\pm39$ & $2542\pm31$ & $2300\pm27$ & $1913\pm23$ & $1265\pm15$ & $478\pm6$ \\\hline
		$|m_t^{\rm rec} - m_t| < 50$ GeV & $2763\pm41$ & $2507\pm36$ & 
$2152\pm29$ & $1961\pm25$ & $1625\pm21$ & $1075\pm14$ & $403\pm5$ \\\hline
		$m_{\rm total} < 1$ TeV & $2713\pm40$ & $2468\pm36$ & 
$2089\pm28$ & $1893\pm25$ & $1556\pm21$ & $1013\pm13$ & $375\pm5$ \\\hline
		$|m_S^{\rm rec} - m_S| < 30$ GeV & $1908\pm34$ & $1729\pm30$ & 
$1440\pm23$ & $1297\pm21$ & $1062\pm17$ & $690\pm11$ & $252\pm4$ \\\hline
	\end{tabular}
	}
	\captionsetup{width=1.4\textwidth}
	\caption{Event yields after each cut for the seven benchmark signal 
points, in the analysis for $t\to S u, S\to\mu^+\mu^-$. \mr{We fix $Y_{13} = Y_{31} = 0.1$, $\Lambda = 1$ TeV and $\mathcal{B}(S\to \mu^+\mu^-) = 1$.} The event yields presented assume a collected luminosity of $\mathcal{L} = 150~{\rm fb}^{-1}$.}
	\label{tab:First_Signal_Up}
\end{table}

\begin{table}[t]
	\centering
	\scalebox{0.75}{
	\begin{tabular}{|l|c|c|c|c|c|c|c|}
		\hline
		Cuts/$m_S$ & 20 GeV & 50 GeV & 80 GeV & 90 GeV & 100 GeV & 120 GeV & 150 GeV\\\hline
		basic & $5235\pm51$ & $4923\pm46$ & $4070\pm35$ & $3633\pm31$ & $3050\pm26$ & $1836\pm15$ & $362\pm3$ \\\hline
		$n_j > 1$ & $5214\pm51$ & $4906\pm46$ & $4050\pm35$ & $3616\pm31$ & $3030\pm26$ & $1821\pm15$ & $352\pm3$ \\\hline
		$n_b = 1$ & $2705\pm37$ & $2520\pm33$ & $2103\pm25$ & $1870\pm22$ & $1571\pm18$ & $957\pm11$ & $188\pm2$ \\\hline
		$n_{\mu^+ \mu^-} = 1$ & $2705\pm37$ & $2520\pm33$ & $2102\pm25$ & $1870\pm22$ & $1571\pm18$ & $957\pm11$ & $188\pm2$ \\\hline
		$|m_t^{\rm rec} - m_t| < 50$ GeV & $2229\pm33$ & $2072\pm30$ & 
$1754\pm23$ & $1551\pm20$ & $1311\pm17$ & $801\pm10$ & $161\pm2$ \\\hline
		$m_{\rm total} < 1$ TeV & $2194\pm33$ & $2038\pm29$ & 
$1708\pm23$ & $1488\pm20$ & $1248\pm16$ & $749\pm10$ & $148\pm2$ \\\hline
		$|m_S^{\rm rec} - m_S| < 30$ GeV & $1502\pm28$ & $1406\pm24$ & 
$1166\pm19$ & $1003\pm16$ & $829\pm13$ & $497\pm8$ & $100\pm2$ \\\hline
	\end{tabular}
	}
	\captionsetup{width=1.4\textwidth}
	\caption{Event yields after each cut for the seven benchmark signal points, in the analysis for $t\to S c, S\to\mu^+\mu^-$. \mr{We fix $Y_{23} = Y_{32} = 0.1$, $\Lambda = 1$ TeV and $\mathcal{B}(S\to \mu^+\mu^-) = 1$.} The event yields presented assume a collected luminosity of $\mathcal{L} = 150~{\rm fb}^{-1}$.
	}
	\label{tab:First_Signal_Charm}
\end{table}

\begin{table}[t]
	\centering
	\scalebox{0.75}{
	\begin{tabular}{|l|c|c|c|c|c|c|c|}
		\hline
		Cuts/Background & $tW$ & $t\bar{t}W/t\bar{t}Z$ & $ZZZ/WWZ$ & $ZZ/WZ/WW$ & $t\bar{t}$ & $tZ$ \\\hline
		basic & $334\pm236$ & $5.2\pm0.7$ & $10.1\pm0.5$ & $16615\pm2220$ & $348\pm24$ & $128\pm2$ \\\hline
		$n_j > 1$ & $334\pm236$ & $5.2\pm0.7$ & $9.1\pm0.5$ & $8011\pm1542$ & $326\pm24$ &  $128\pm2$ \\\hline
		$n_b = 1$ & $334\pm236$ & $2.2\pm0.5$ & $1.0\pm0.2$ & $<~74$ & $172\pm17$ & $65\pm1$ \\\hline
		$n_{\mu^+ \mu^-} = 1$ & $<~42$ & $1.1\pm0.3$ & $0.7\pm0.1$ & --- & $172\pm17$ & $39\pm1$ \\\hline
		$|m_t^{rec} - m_t| < 50$ GeV & --- & $0.7\pm0.3$ & $0.4\pm0.1$ & --- & $114\pm14$ & $31\pm1$ \\\hline
		$m_{total} < 1$ TeV & --- & $0.3\pm0.2$ & $0.25\pm0.08$ & --- & $80\pm12$  & $23.2\pm0.9$ \\\hline
	\end{tabular}
	}
	\captionsetup{width=1.4\textwidth}
	\caption{Event yields after each cut for the dominant backgrounds, 
in the analysis for $t\to Sq, S\to\mu^+\mu^-$. 
The $Z$ + jets sample is reduced to negligible values after the cut on 
the lepton multiplicity. The event yields presented assume a collected 
luminosity of $\mathcal{L} = 150~{\rm fb}^{-1}$.}
	\label{tab:First_Background}
\end{table}
\end{landscape}

\begin{table}[t]
	\centering
	\scalebox{0.8}{
	\begin{tabular}{|l|c|c|c|c|c|c|c|}
		\hline
		Background/$m_S$ & 20 GeV & 50 GeV & 80 GeV & 90 GeV & 100 GeV & 120 GeV & 150 GeV\\\hline
		$t\bar{t}W/t\bar{t}Z$ & $<~0.09$ & $<~0.09$ & $0.2\pm0.1$ & $0.2\pm0.1$ & $0.2\pm0.1$ & $0.2\pm0.1$ & $<~0.09$ \\\hline
		$ZZZ/WWZ$ & $0\pm0$ & $0.06\pm0.04$ & $0.22\pm0.08$ & $0.17\pm0.07$ & $0.17\pm0.07$ & $<~0.03$ & $<~0.03$ \\\hline
		$t\bar{t}$ ($\mu$) & $12\pm4$ & $34\pm8$ & $39\pm8$ & $39\pm8$ & $36\pm8$ & $15\pm5$ & $3\pm2$ \\\hline
		$tZ$ & $1.8\pm0.2$ & $3.3\pm0.3$ & $18.6\pm0.8$ & $18.4\pm0.8$ & $17.9\pm0.8$ & $11.4\pm0.6$ & $0.8\pm0.2$ \\\hline
	\end{tabular}
	}
	\caption{Event yields for the last selection cut on $m_S^{\rm rec}$ for the dominant backgrounds, in the analysis for $t\to Sq, S\to\mu^+\mu^-$. The event yields presented assume a collected luminosity of $\mathcal{L} = 150~{\rm fb}^{-1}$.}
	\label{tab:First_Background_MassCut}
\end{table}
\section{Search for $t\to Sq, S\to\tau^+\tau^-$}
\label{sec:tStt}

We focus now on the scenario where the scalar $S$ decays to a 
pair of taus, concentrating on the hadronic decays of the latter. Again, we 
focus on the leptonic decay of the $W$. Thus, we require events to contain 
exactly one light lepton and at least three jets, from which exactly one must be 
$b$-tagged and exactly two must be tagged as taus decaying into hadrons. The 
efficiency for $\tau$-tagging is $60\%$ 
while
the misidentification rate is $1\%$ with no dependency on the 
transverse momentum. 
Jets and leptons are defined in the same $p_T$ and $|\eta|$ ranges as in the 
previous analysis; the effect of these requirements in conjunction with the cuts on the number of leptons and hadronic taus can be found in the yields tables, labeled as ``basic''.

The dominant backgrounds for this channel are the exclusive $tW$ and 
$t\overline{t}$ processes, where the top quark is assumed to decay to $Wb$ and 
$W\rightarrow \tau \nu$. Indeed, after the aforementioned cuts on the 
particle multiplicities, the other background components (with the largest 
cross sections) become irrelevant; see Tab.~\ref{tab:Second_Background}.

\begin{figure}[t]
 \centering
  \includegraphics[width=0.49\textwidth]{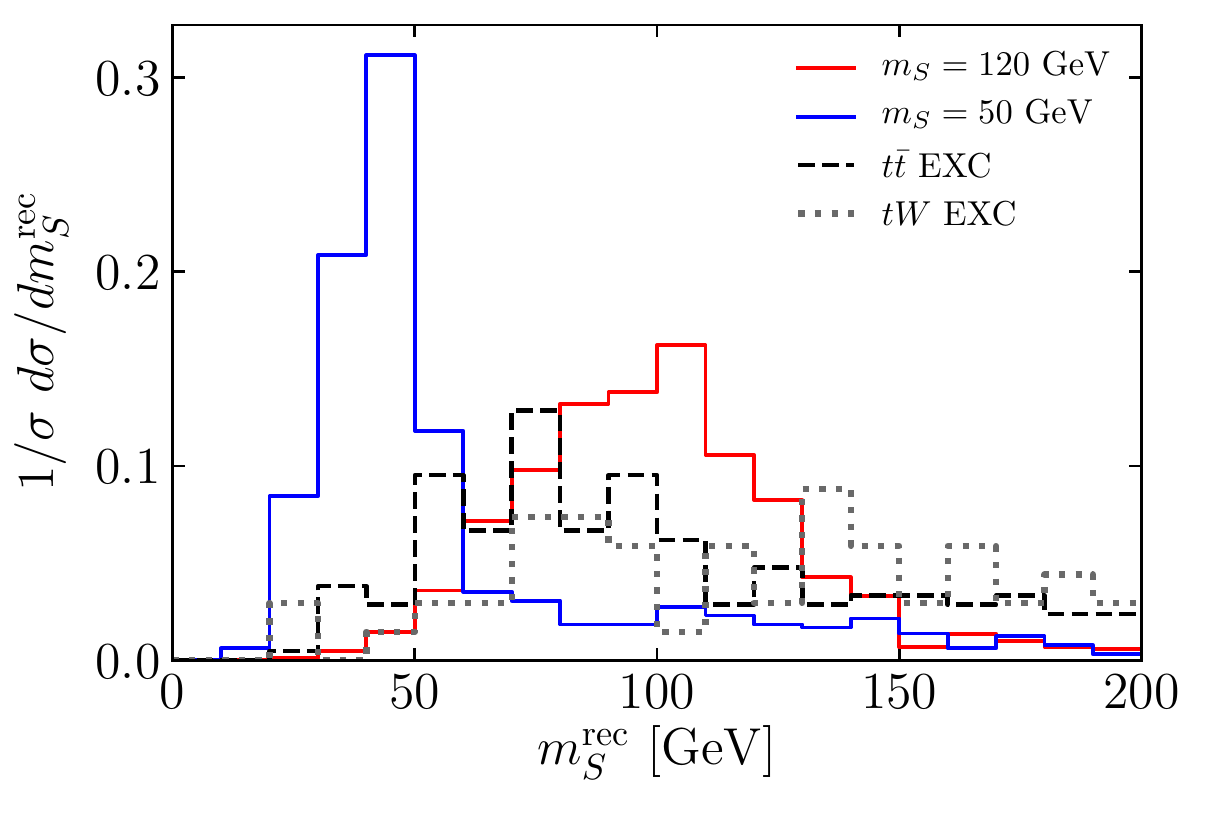}
  \includegraphics[width=0.5\textwidth]{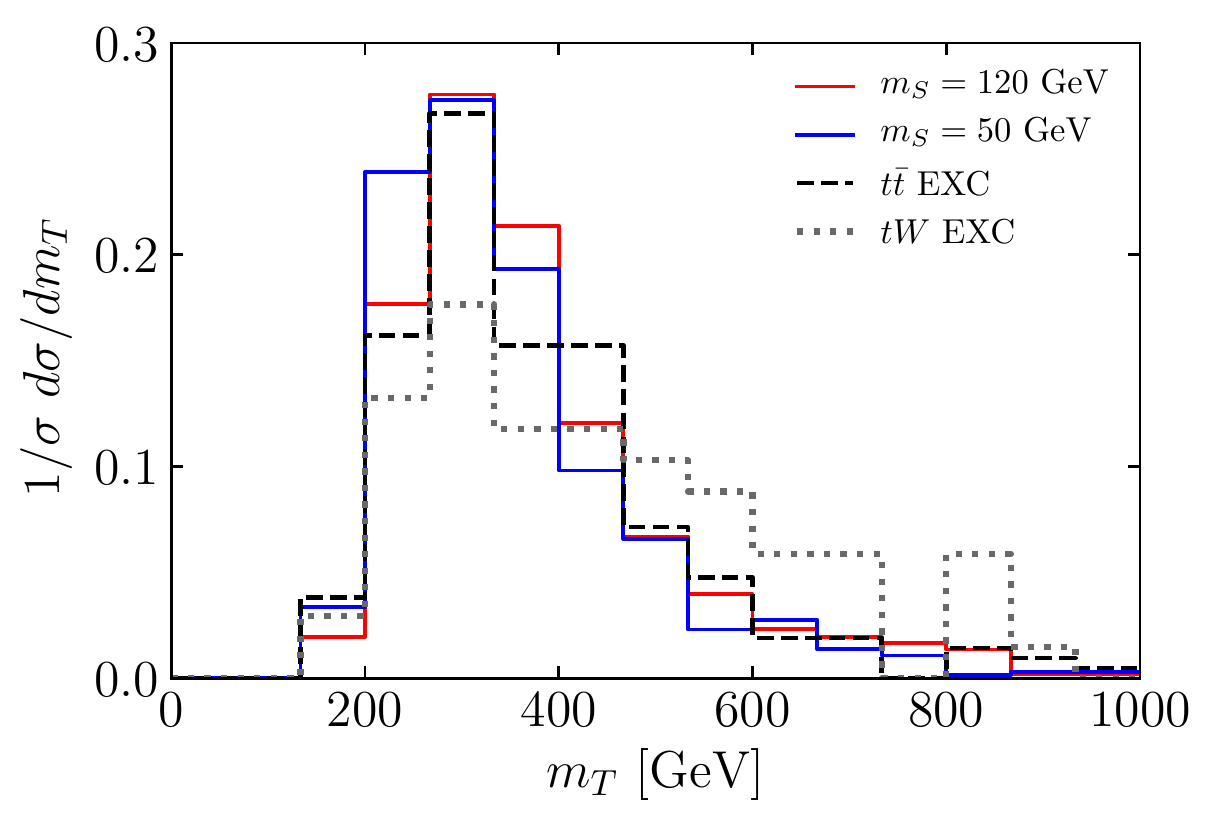}\\
    \includegraphics[width=0.49\textwidth]{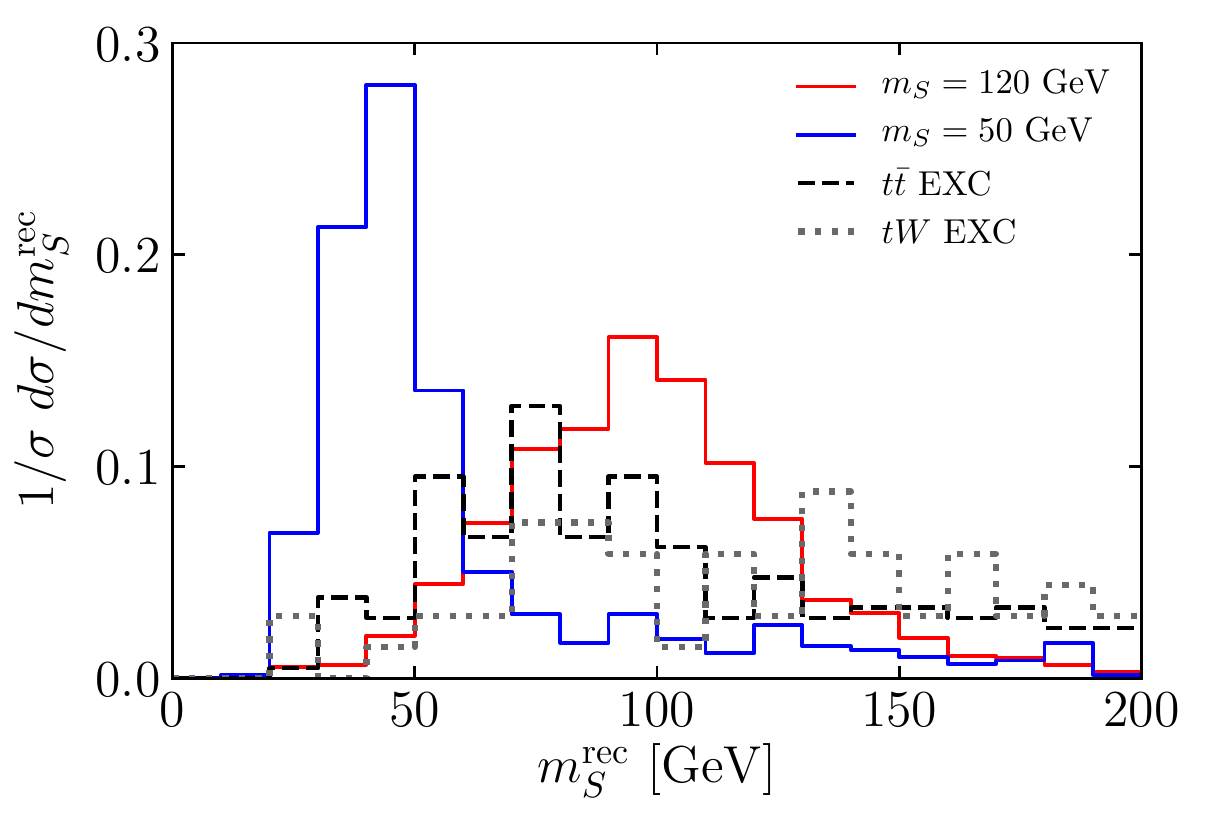}
  \includegraphics[width=0.5\textwidth]{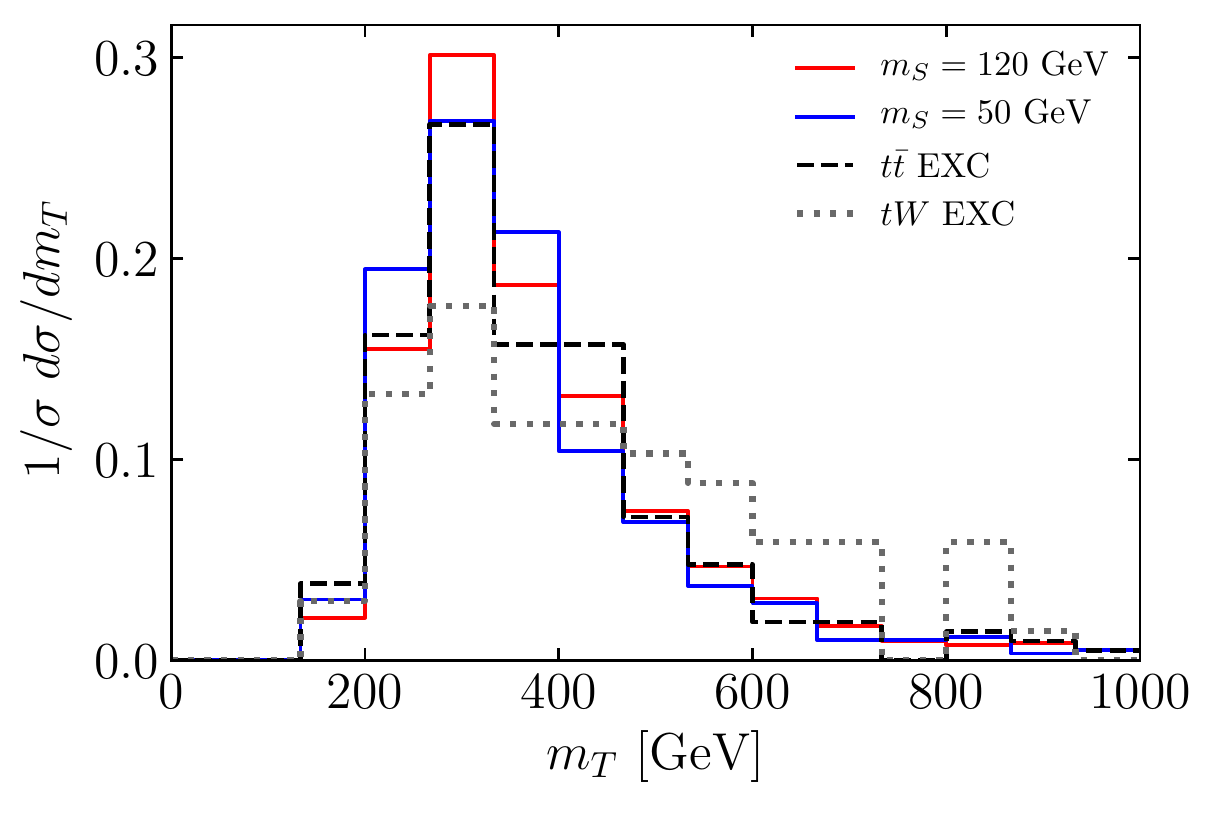}\\
 \caption{The reconstructed scalar mass (left) and the transverse mass  
of the system composed by the lepton, the $b$-jet,
the reconstructed $S$ and missing energy (right) in the analysis proposed for 
$t\to Sq$, $S\to\tau^+\tau^-$. In the upper (bottom) panel, $q = c~(u)$. 
We represent 
the distributions of two signal benchmark points and the two major background 
components, after the cut on the particle multiplicities; the background 
samples are generated exclusively, $i.e.$ only gauge boson decays into taus 
are included. The distributions assume a collected luminosity of $\mathcal{L} 
= 150~{\rm fb}^{-1}$.}
 \label{fig:2nd_analysis}
\end{figure}

We (partially) reconstruct the scalar $S$ from the two tau-jets obtaining its invariant mass, $m_S^{\rm rec}$. In the left panels of Fig.~\ref{fig:2nd_analysis} we 
show the normalized distribution of this variable, after the basic selection 
cuts, in two signal BPs and in the main backgrounds. In the same figure, we 
plot the transverse mass distribution of the system composed by the lepton, the 
$b$-tagged jet, the reconstructed scalar and the missing energy. We require 
this latter variable to be smaller than 500 GeV. Finally, events are required to fulfill $|m_S - m_S^{rec}| < 30$~GeV, where $m_S^{rec}$ is the mass of the reconstructed scalar candidate and $m_S$ is the corresponding value being probed.

The cut flows for the signal in the up quark and in the
 charm quark cases are given in Tabs.~\ref{tab:Second_Signal_Up} and
\ref{tab:Second_Signal_Charm}; respectively. Likewise, 
Tab.~\ref{tab:Second_Background} shows the $m_S$-independent yields for the
different backgrounds and in Tab.~\ref{tab:Second_Background_MassCut} we write 
the mass-dependent ones.
Similarly to the previous analysis, the expected upper limit on the signal 
strength, $\sigma_{95\%}/\sigma_{\rm th}$ ($pp \to tS(q)$, $S \to \tau^+\tau^-$), is obtained using the invariant mass of the scalar $S$ candidate 
distributed into 20 bins.
The 95\% CL upper limits on the branching ratio $\mathcal{B}\left(t\rightarrow S q, S\rightarrow 
\tau^+\tau^-\right)$ and on the cross section $\sigma (p p \to t S (q), S 
\to \tau^+\tau^-)$ are shown in Fig.~\ref{fig:BR_tautau}. 
\begin{figure}[!ht]
 \centering
  \includegraphics[width=0.49\textwidth]{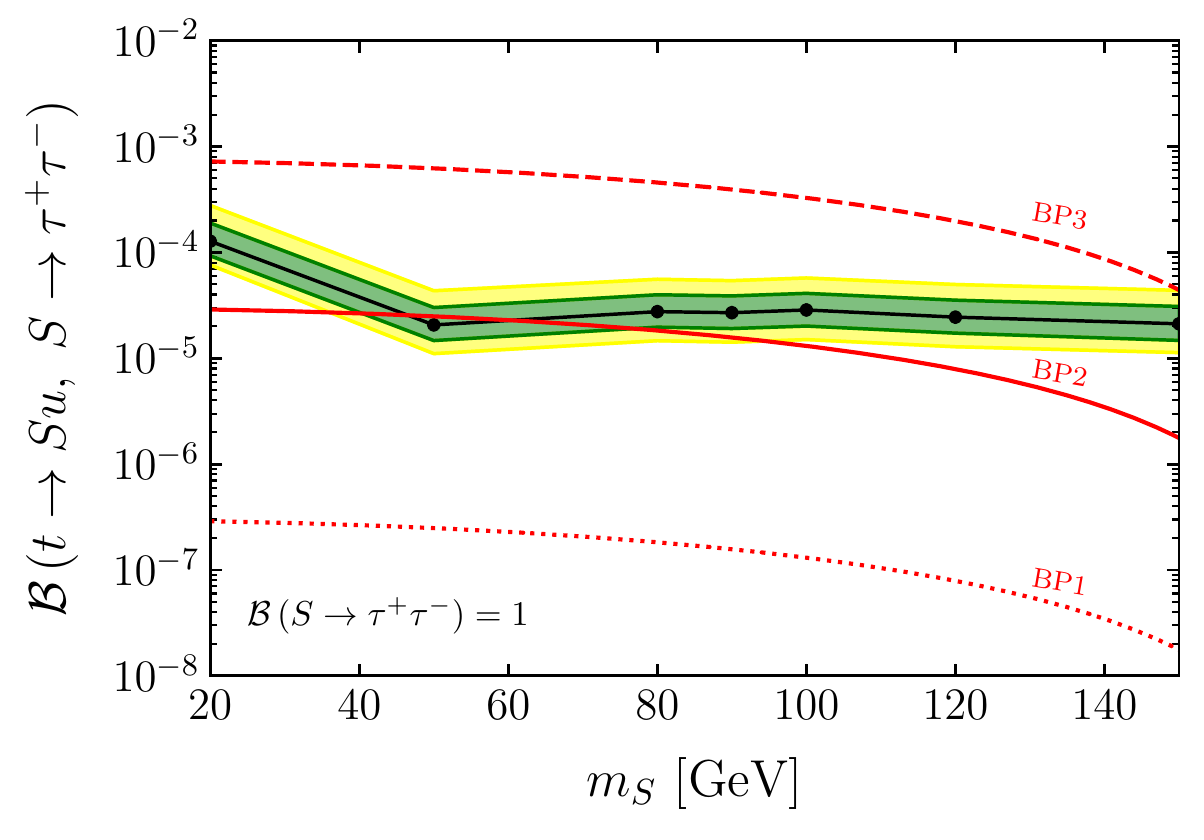}
  \includegraphics[width=0.49\textwidth]{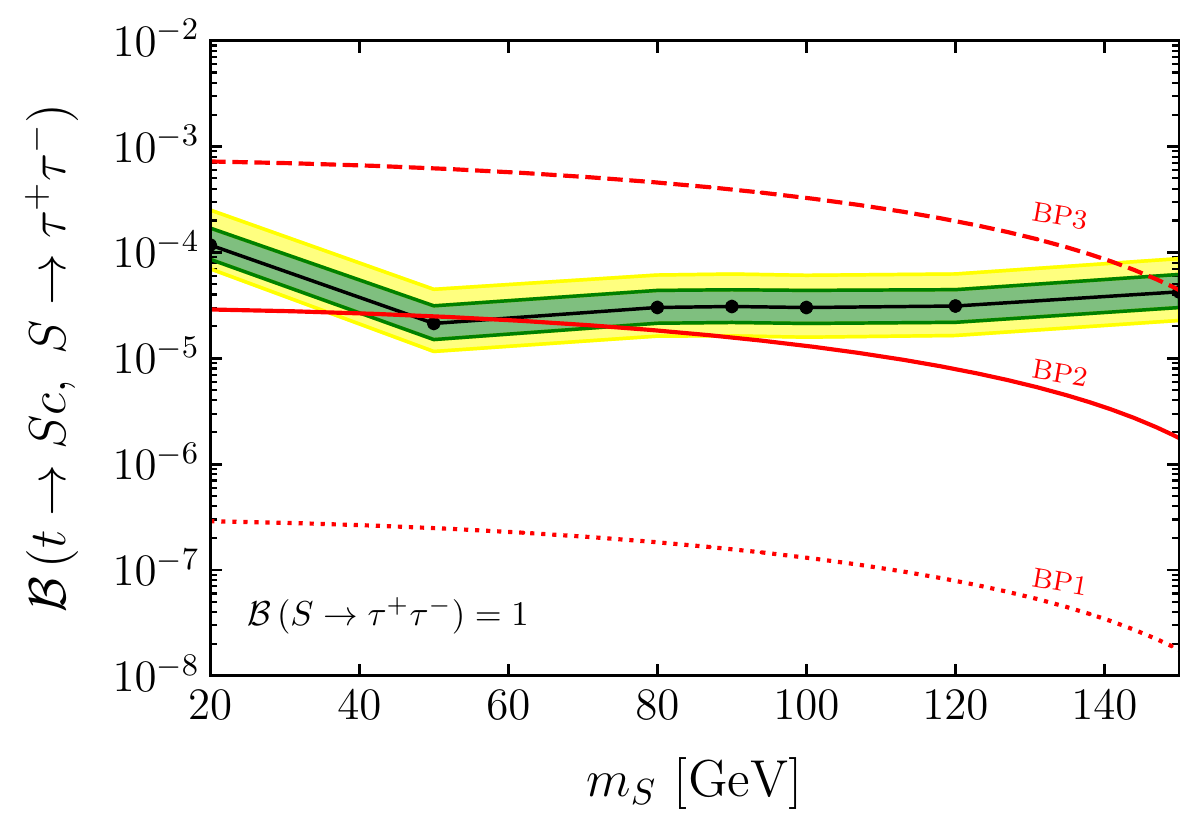}\\
      \includegraphics[width=0.49\textwidth]{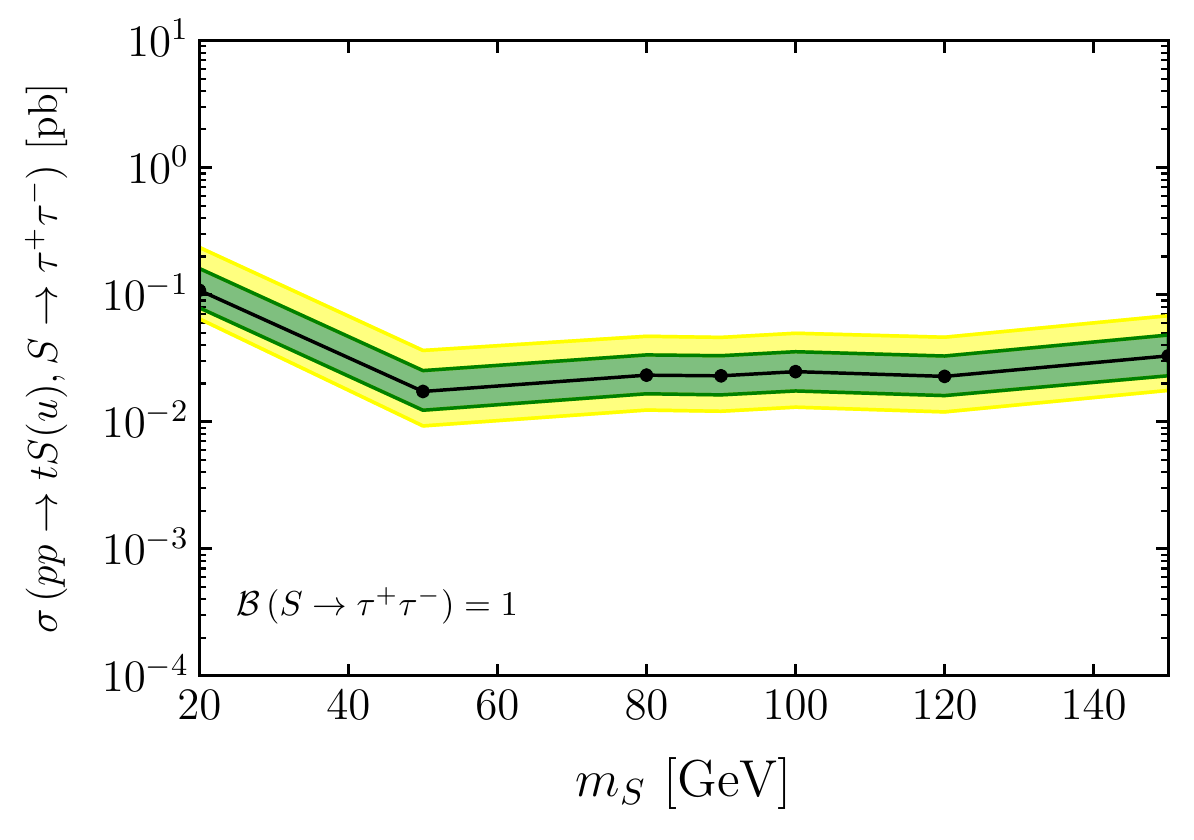}
  \includegraphics[width=0.49\textwidth]{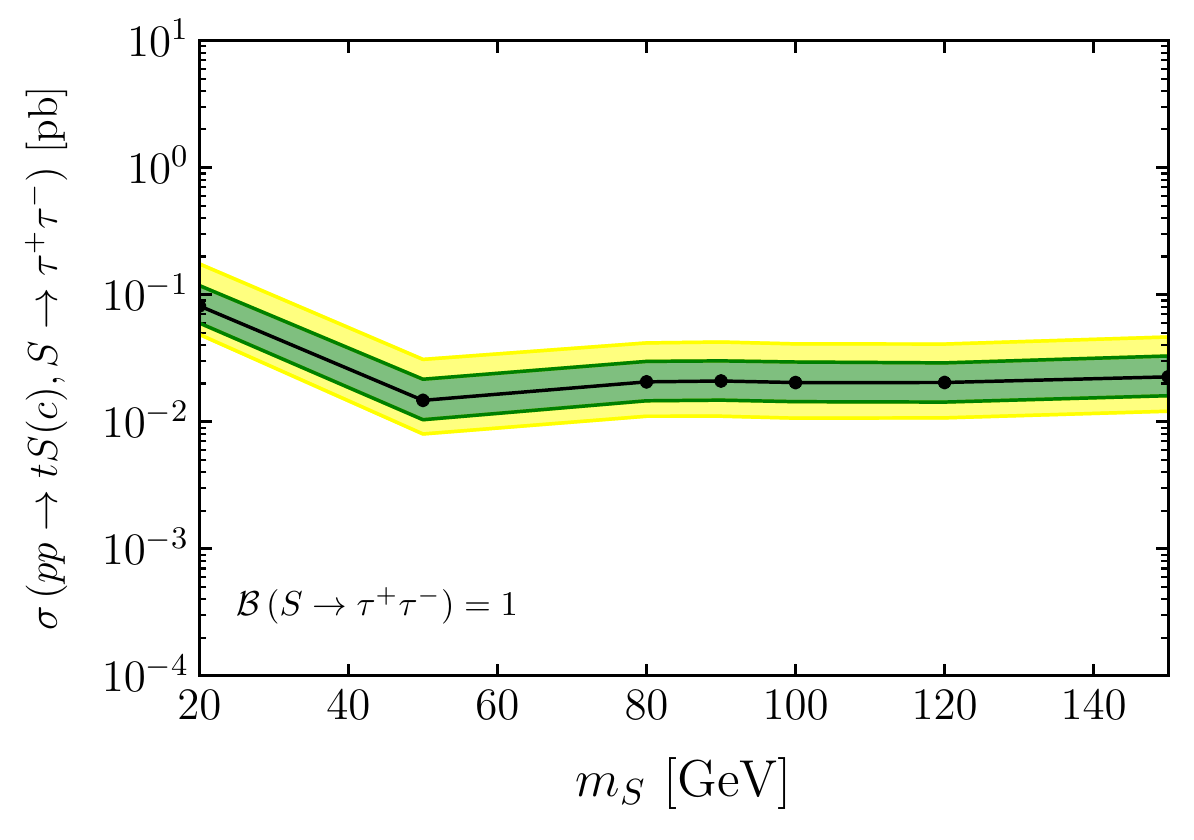}\\
 \caption{In the upper (bottom) panels, we show the 95\% CL limits on the 
branching ratio (cross section times branching ratio) that can be tested in the 
$\tau^+\tau^-$ channel, in the 
analysis proposed for $t\rightarrow S q,~S\rightarrow \tau^+\tau^-$, with $q = 
u~(c)$ in the panels on the left (right).  The green and yellow bands 
show the $\pm 1 \sigma$ and $\pm 2\sigma$ uncertainty on the limits, 
respectively. The limits are obtained for a 
collected luminosity $\mathcal{L} = 150~{\rm fb}^{-1}$. Superimposed are the 
theoretical expectations in three BPs. }
 \label{fig:BR_tautau}
\end{figure}
\begin{landscape}
\begin{table}[t]
	\centering
	\scalebox{0.75}{
	\begin{tabular}{|l|c|c|c|c|c|c|c|}
		\hline
		Cuts/$m_S$ & 20 GeV & 50 GeV & 80 GeV & 90 GeV & 100 GeV & 120 GeV & 150 GeV\\\hline
		basic & $267\pm13$ & $596\pm17$ & $586\pm15$ & $561\pm14$ & $469\pm11$ & $339\pm8$ & $131\pm3$ \\\hline
		$n_j > 3$ & $264\pm13$ & $588\pm17$ & $578\pm15$ & $553\pm13$ & $461\pm11$ & $331\pm8$ & $126\pm3$ \\\hline
		$n_b = 1$ & $118\pm8$ & $306\pm12$ & $309\pm11$ & $330\pm10$ & $284\pm8$ & $171\pm6$ & $66\pm2$ \\\hline
		$m_T < 500$ GeV & $86\pm7$ & $259\pm11$ & $256\pm910$ & $280\pm10$ & $234\pm8$ & $143\pm5$ & $47\pm2$ \\\hline
		$|m_S^{rec} - m_S| < 30$ GeV & $47\pm5$ & $220\pm11$ & $193\pm9$ & $236\pm9$ & $195\pm7$  & $79\pm4$ & $18\pm1$ \\\hline
	\end{tabular}
	}
	\captionsetup{width=1.4\textwidth}
	\caption{Event yields after each cut for the seven benchmark signal points, in the analysis for $t\to S u, S\to\tau^+\tau^-$. \mr{We use $Y_{13} = Y_{31} = 0.1$, $\Lambda = 1$ TeV and $\mathcal{B}(S\to \tau^+\tau^-) = 1$}. The event yields presented assume a collected luminosity of $\mathcal{L} = 150~{\rm fb}^{-1}$.}
	\label{tab:Second_Signal_Up}
\end{table}
\begin{table}[t]
	\centering
	\scalebox{0.75}{
	\begin{tabular}{|l|c|c|c|c|c|c|c|}
		\hline
		Cuts/$m_S$ & 20 GeV & 50 GeV & 80 GeV & 90 GeV & 100 GeV & 120 GeV & 150 GeV\\\hline
		basic & $221\pm10$ & $540\pm15$ & $486\pm12$ & $462\pm11$ & $410\pm9$ & $263\pm6$ & $54\pm1$ \\\hline
		$n_j > 3$ & $219\pm10$ & $536\pm15$ & $482\pm12$ & $458\pm11$ & $407\pm9$ & $260\pm6$ & $52\pm1$ \\\hline
		$n_b = 1$ & $107\pm7$ & $276\pm11$ & $256\pm9$ & $297\pm8$ & $215\pm7$ & $131\pm4$ & $27.5\pm0.8$ \\\hline
		$m_T < 500$ GeV & $91\pm7$ & $243\pm10$ & $222\pm8$ & $244\pm8$ & $182\pm6$ & $111\pm4$ & $22.9\pm0.7$ \\\hline
		$|m_S^{rec} - m_S| < 30$ GeV & $48\pm5$ & $203\pm9$ & $171\pm7$ & $207\pm7$ & $124\pm5$ & $62\pm3$ & $9.0\pm0.5$ \\\hline
	\end{tabular}
	}
	\captionsetup{width=1.4\textwidth}
	\caption{Event yields after each cut for the seven benchmark signal 
points, 
in the analysis for $t\to S c, S\to\tau^+\tau^-$. \mr{We fix $Y_{23} = Y_{32} = 0.1$, $\Lambda = 1$ TeV and $\mathcal{B}(S\to \tau^+\tau^-) = 1$.} The event yields presented assume a collected luminosity of $\mathcal{L} = 150~{\rm fb}^{-1}$.}
	\label{tab:Second_Signal_Charm}
\end{table}

\begin{table}[t]
	\centering
	\scalebox{0.75}{
	\begin{tabular}{|l|c|c|c|c|c|c|c|}
		\hline
		Cuts/Background & $tW$ ($\tau$) & $t\bar{t}W/t\bar{t}Z$ & $ZZZ/WWZ$ & $ZZ/WZ/WW$ & $t\bar{t}$ ($\tau$) & $tZ$ \\\hline
		basic & $48\pm4$ & $1.0\pm0.3$ & $1.4\pm0.2$ & $297\pm133$ & $99\pm5$ & $1.3\pm0.2$ \\\hline
		$n_j > 3$ & $44\pm3$ & $1.0\pm0.3$ & $1.2\pm0.2$ & $178\pm103$ & $96\pm5$ & $1.3\pm0.2$ \\\hline
		$n_b = 1$ & $19\pm2$ & $0.6\pm0.2$ & $0.06\pm0.04$ & $<~74$ & $52\pm4$ & $0.4\pm0.1$ \\\hline
		$m_T < 500$ GeV & $11\pm2$ & $0.4\pm0.2$ & $0.05\pm0.04$ & --- & $42\pm3$ & $0.3\pm0.1$ \\\hline
	\end{tabular}
	}
	\captionsetup{width=1.4\textwidth}
	\caption{Event yields after each cut for the dominant backgrounds, in 
the analysis for $t\to Sq, S\to\tau^+\tau^-$.  
The $Z$ + jets sample is reduced to negligible values after the fifth cut. 
The event yields presented assume a collected luminosity of $\mathcal{L} = 
150~{\rm fb}^{-1}$.}
	\label{tab:Second_Background}
\end{table}
\end{landscape}
\begin{table}[t]
	\centering
	\scalebox{0.76}{
	\begin{tabular}{|l|c|c|c|c|c|c|c|}
		\hline
		Background/$m_S$ & 20 GeV & 50 GeV & 80 GeV & 90 GeV & 100 GeV & 120 GeV & 150 GeV\\\hline
		$tW$ ($\tau$)& $0.6\pm0.4$ & $3.1\pm0.9$ & $5\pm1$ & $5\pm1$ & $5\pm1$ & $5\pm1$ & $4\pm1$ \\\hline
		$t\bar{t}W/t\bar{t}Z$ & $<~0.09$ & $<~0.09$ & $0.3\pm0.2$ & $0.3\pm0.2$ & $0.2\pm0.1$ & $<~0.09$ & $<~0.09$ \\\hline
		$ZZZ/WWZ$ & $<~0.03$ & $<~0.03$ & $<~0.03$ & $<~0.03$ & $<~0.03$ & $<~0.03$ & $<~0.03$ \\\hline
		$t\bar{t}$ ($\tau$) & $3.7\pm0.9$ & $17\pm2$ & $24\pm2$ & $21\pm2$ & $20\pm2$ & $14\pm2$ &$10\pm1$  \\\hline
		$tZ$ & $<~0.03$ & $0.07\pm0.05$ & $0.10\pm0.06$ & $0.10\pm0.06$ & $0.10\pm0.06$ & $0.10\pm0.06$ & $0.14\pm0.07$ \\\hline
	\end{tabular}
	}
	\caption{Event yields for the last selection cut on $m_S^{\rm rec}$ 
for the dominant 
backgrounds, 
in the analysis for $t\to Sq, S\to\tau^+\tau^-$. The event yields presented 
assume a collected 
luminosity of $\mathcal{L} = 150~{\rm fb}^{-1}$.  }
	\label{tab:Second_Background_MassCut}
\end{table}
\begin{figure}[ht]
 \centering
  \includegraphics[width=0.49\textwidth]{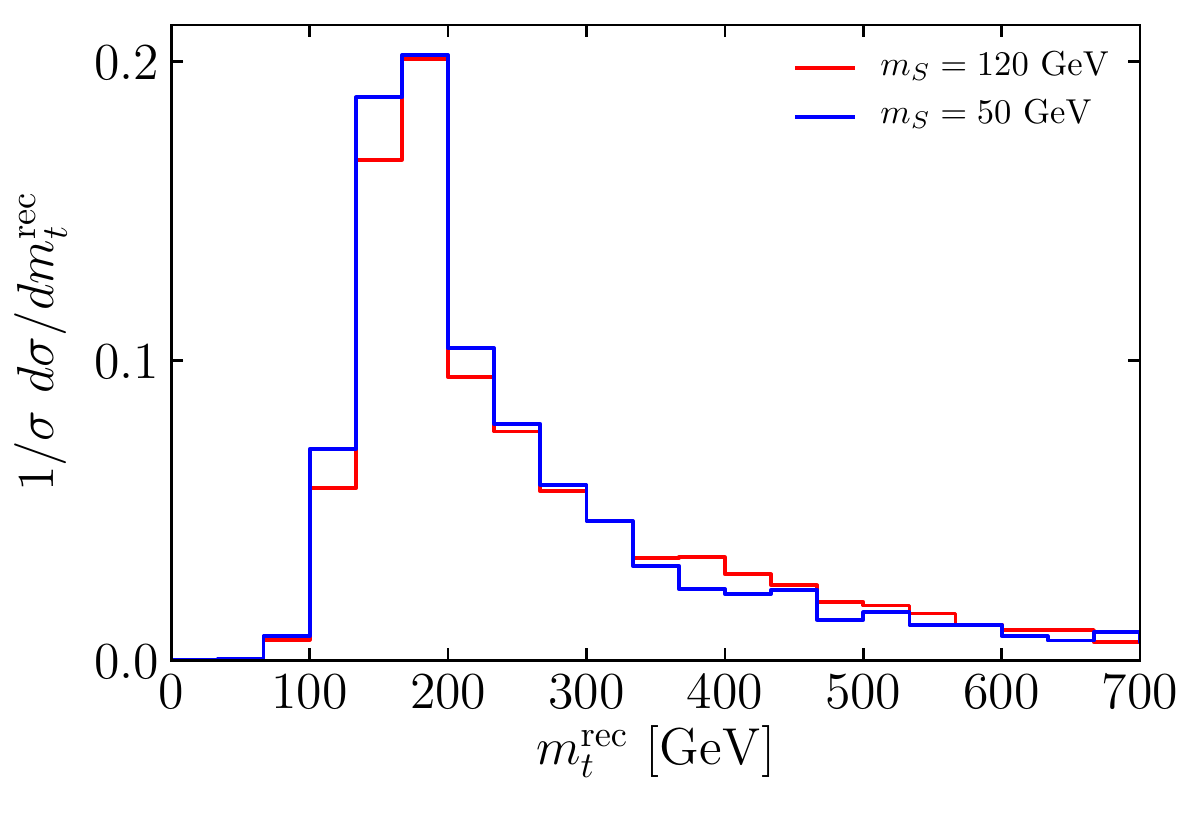}
   \includegraphics[width=0.49\textwidth]{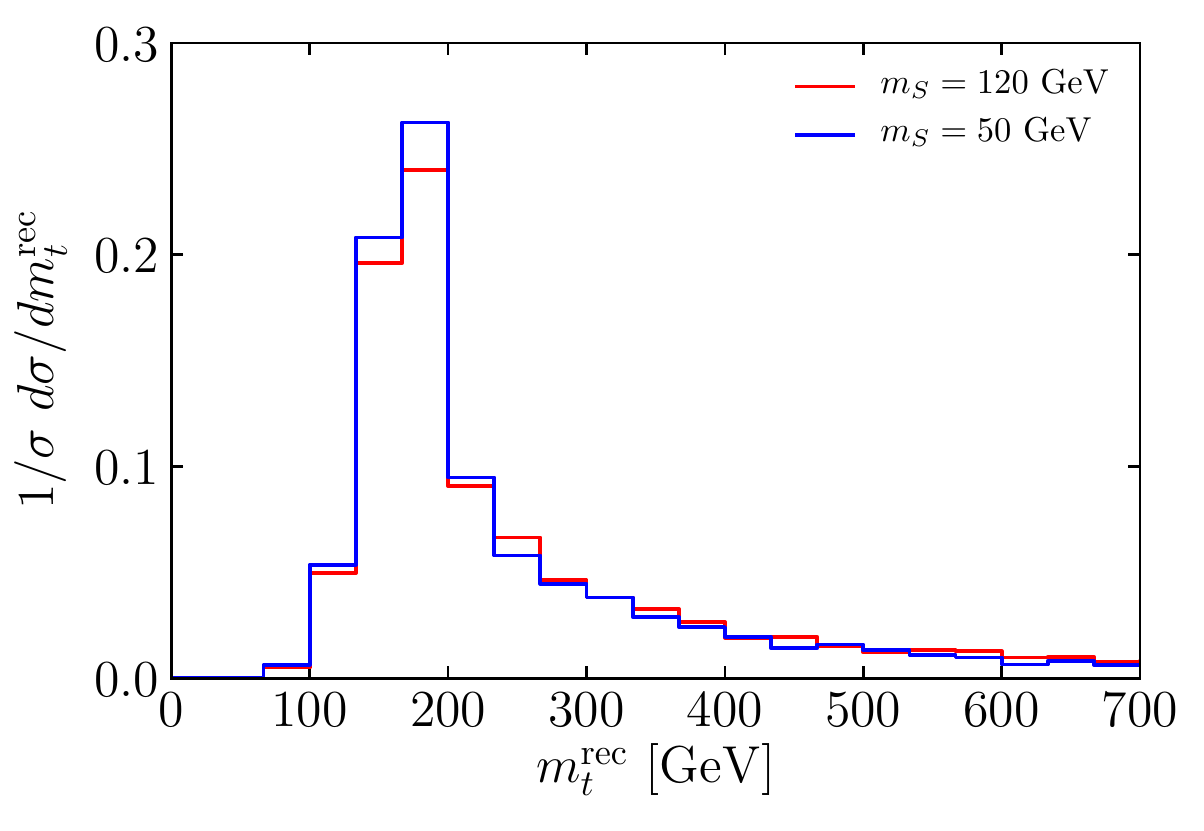}\\
 \caption{The reconstructed top mass in the analysis proposed for $t\rightarrow S S q,~S\rightarrow \mu^+\mu^-$. In the left (right) panel, $q = c~(u)$. The distributions assume a collected luminosity of $\mathcal{L} = 150~{\rm fb}^{-1}$.}
 \label{fig:3rd_top}
 \end{figure}

\section{Search for $t\to SSq, S\to\mu^+\mu^-$}
\label{sec:tSSll}

Finally, let us
 develop an analysis to search for two singlet scalars produced in association 
with a top quark (or in the decay of a top quark in pair production), 
both decaying to a pair of muons. \mr{We focus on the hadronic decays of the 
$W$.} The final state consists 
of four isolated 
leptons
and at least three jets, one of them required to 
be $b$-tagged. 
Jets and leptons are defined in the same $p_T$ and $|\eta|$ ranges as before; the effect of these requirements, together with the cut on the lepton multiplicity, can be found in the yields tables labeled as ``basic''.
Due to 
the large lepton multiplicity, all the background components are significantly 
reduced; with the additional cuts on the number of jets, most become negligible. 
We are left with six background events from the $t\bar{t}\mr{V}$ and $t\overline{t}$ 
exclusive samples; see Tab.~\ref{tab:Third_Background}.

We reconstruct the top quark from a $W$ boson and a $b$-jet; its invariant mass 
being $m_t^{\rm rec}$. The $W$ is reconstructed from the two light jets with 
invariant mass closest to $m_W$. We then require $m_t^{\rm rec}$ to be within a 
window of $50$ GeV around the top mass. The two scalars $S$ candidates are reconstructed by requiring two muons with opposite sign, with the event being rejected if no such candidates are found.
The opposite-sign muons reconstructing the two scalars are those minimizing $|m_{S_1}^{\rm 
rec} - m_{S_2}^{\rm rec}|$, with $m_{S_{1,2}}^{\rm rec}$ being the invariant 
mass of each pair of opposite sign muons.
It is also required that the invariant mass of the total system, composed of the 
reconstructed top quark and the two scalars, is smaller than 1 TeV.
We finally request the $m_{S_{1,2}}^{\rm rec}$ to be within a window of $30$ GeV from the mass of $S$ being probed.
\begin{figure}[t]
 \centering
  \includegraphics[width=0.49\textwidth]{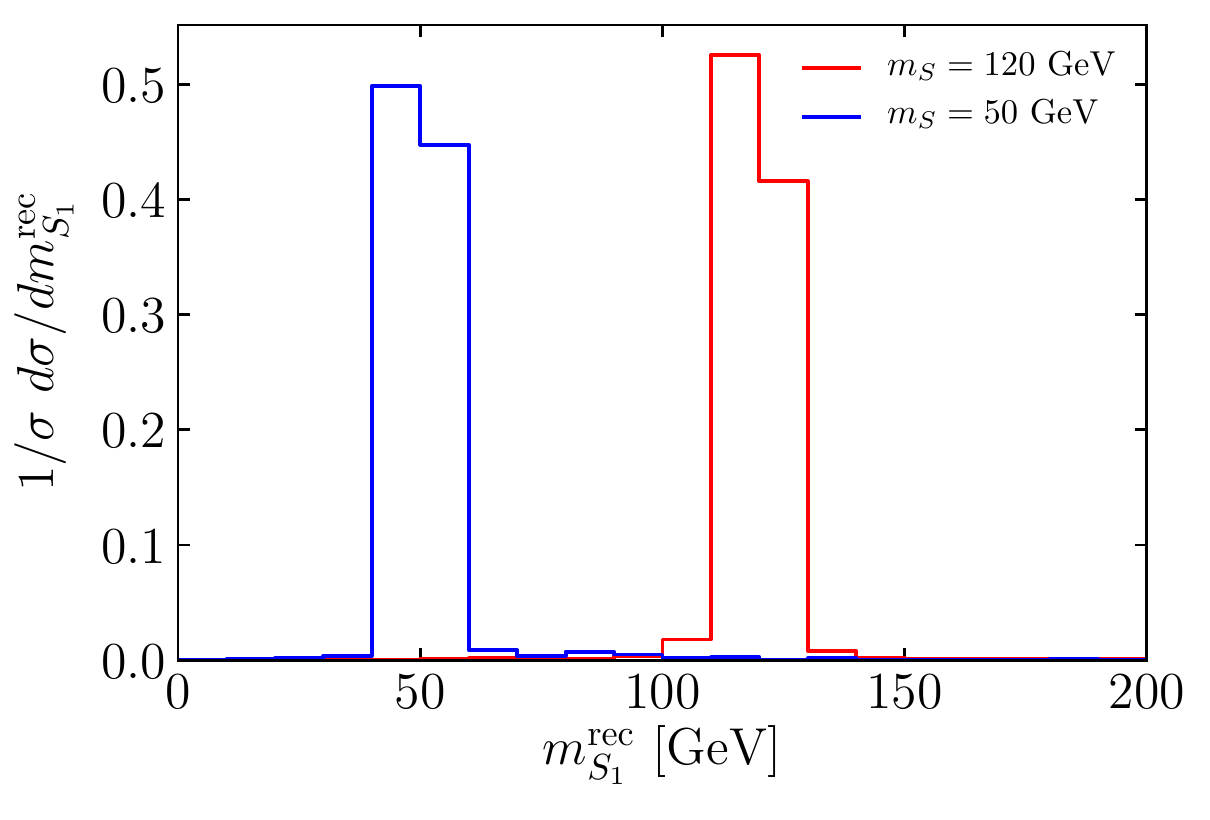}
   \includegraphics[width=0.5\textwidth]{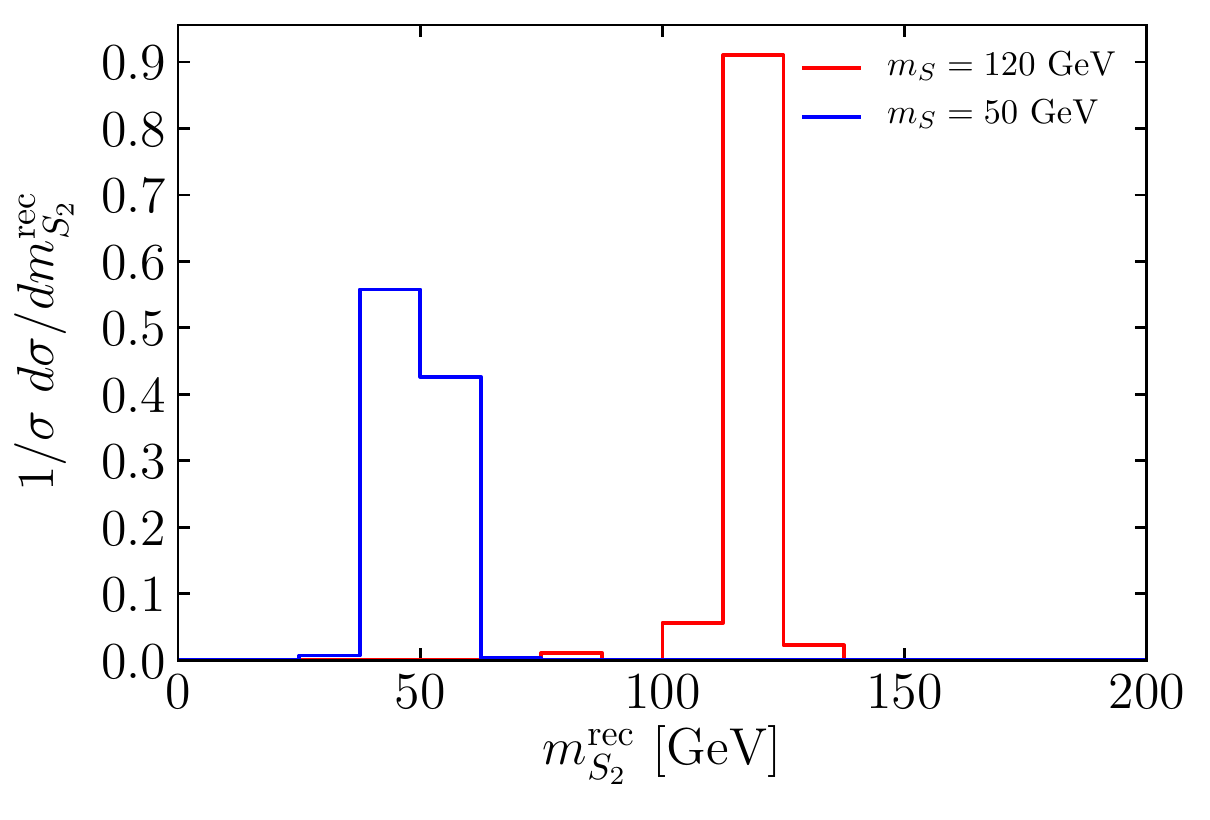}\\
  \includegraphics[width=0.49\textwidth]{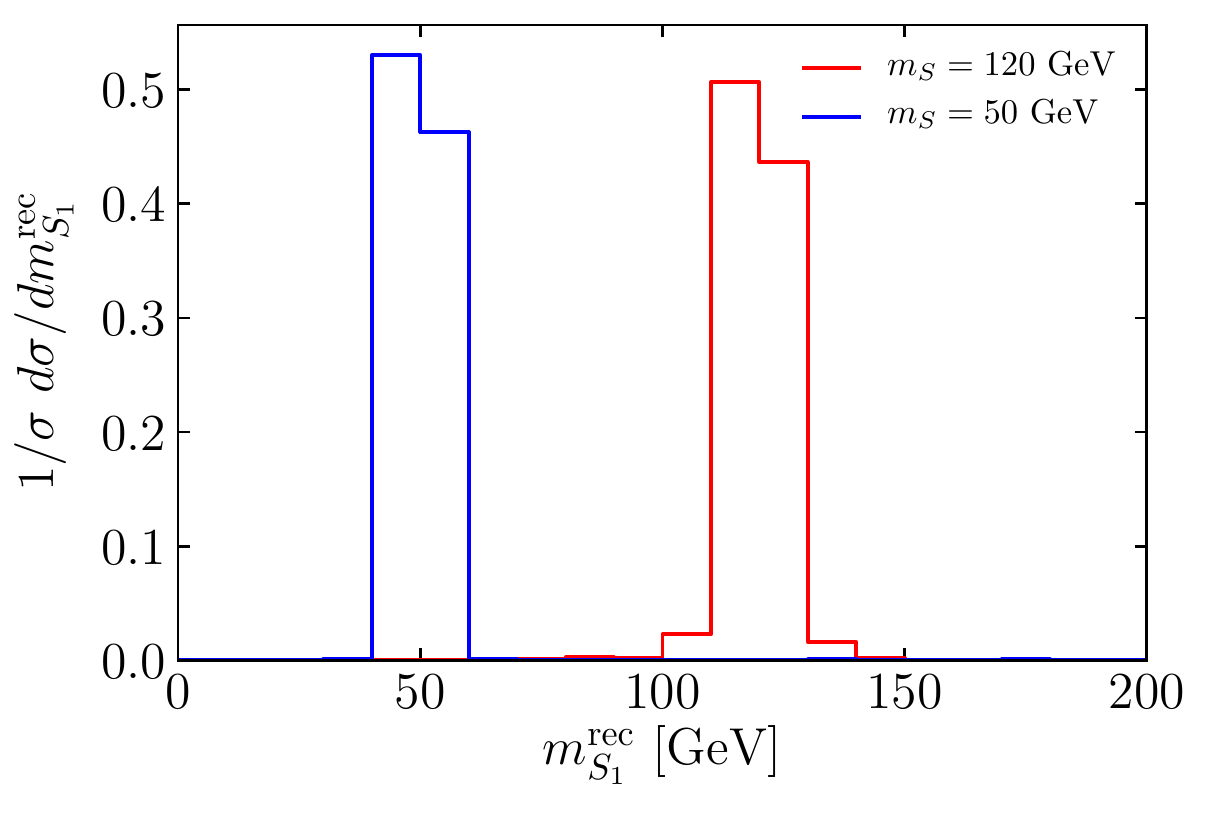}
  \includegraphics[width=0.5\textwidth]{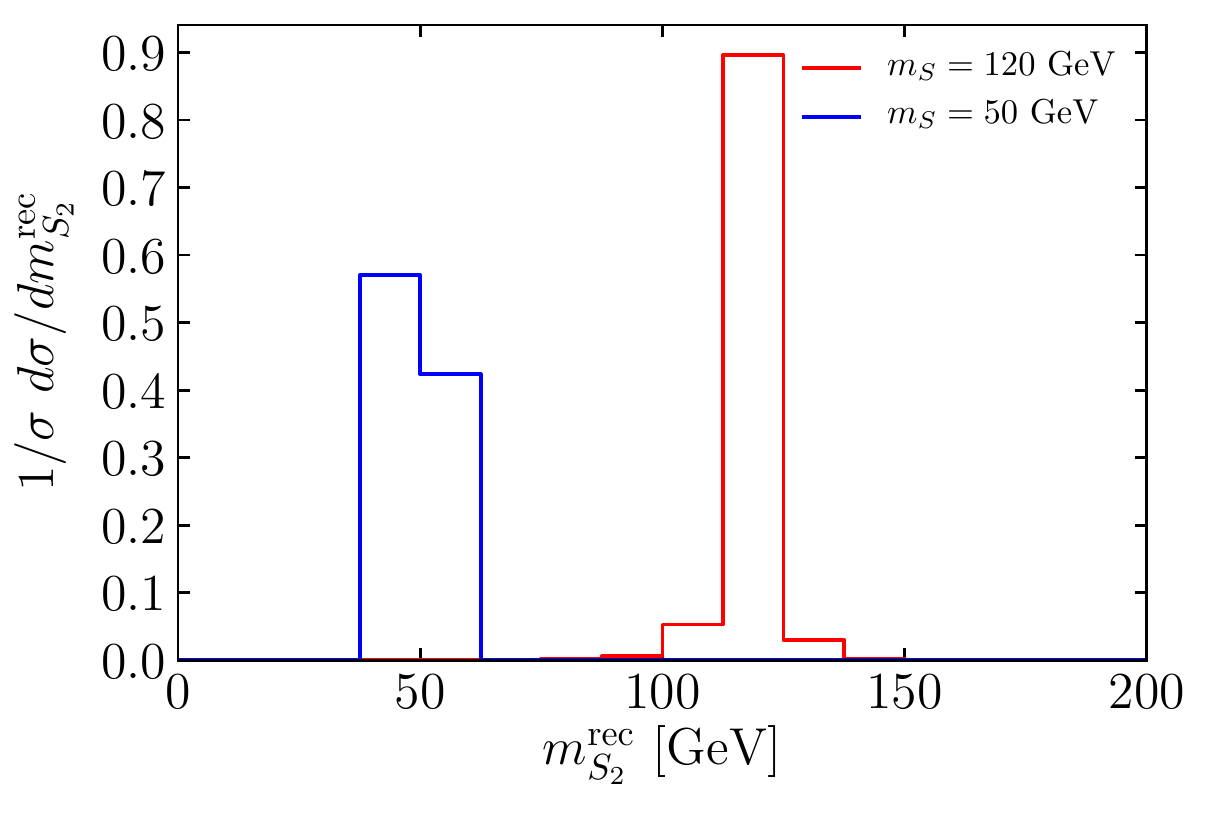}
 \caption{The two reconstructed scalar masses, $m_{S_1}^{\rm rec}$ and 
$m_{S_2}^{\rm rec}$, in the analysis proposed for $t\rightarrow S S 
q,~S\rightarrow \mu^+\mu^-$. In the upper (bottom) panel, $q = c~(u)$. The distributions assume a collected luminosity of $\mathcal{L} = 150~{\rm fb}^{-1}$.}
 \label{fig:3rd_scalar}
 \end{figure}

 In Figs.~\ref{fig:3rd_top} and \ref{fig:3rd_scalar}, we show the normalized 
 distributions of \mr{$m_{t}^{\rm rec}$ and each $m_{S}^{\rm rec}$}, respectively, 
for two signal benchmark points after the basic selection cuts. 

 The cut flows for the signal are given in 
Tabs.~\ref{tab:Third_Signal_Up} and
\ref{tab:Third_Signal_Charm}; in Tab.~\ref{tab:Third_Background} we show the 
scalar mass-independent cut flow for the relevant backgrounds. With the final cut 
on $m_{S_{1,2}}^{\rm rec}$,  
the analysis becomes essentially background-free. 
As before, expected upper limits} on the signal 
strength, $\sigma_{95\%}/\sigma_{\rm th}$ ($pp \to tSS(q)$, $S \to \mu^+\mu^-$), are obtained using the invariant mass of the scalar $S$ candidate 
distributed into 20 bins.
The 95\% upper limits on the branching ratio $\mathcal{B}\left(t\rightarrow S S q, S\rightarrow 
\mu^+\mu^-\right)$ and cross section $\sigma (p p \to t S  S (q), S \to 
\mu^+\mu^-)$ are shown in Fig.~\ref{fig:BR_3mumu}, \mc{again including the $\pm 
1\sigma$ (green band) and the $\pm 2\sigma$ (yellow band) uncertainties.}

\begin{landscape}
\begin{table}[t]
	\centering
	\scalebox{0.64}{
	\begin{tabular}{|l|c|c|c|c|c|c|c|c|c|c|c|}
		\hline
		Cuts/$m_S$ & 20 GeV & 30 GeV & 40 GeV & 50 GeV & 60 GeV & 70 GeV & 80 GeV & 90 GeV & 100 GeV & 120 GeV & 150 GeV \\\hline
		basic & $2.87\pm0.03$ & $2.70\pm0.03$ & $2.58\pm0.03$ & $2.47\pm0.02$ & $2.36\pm0.02$ & $2.27\pm0.02$ & $2.25\pm0.02$ & $1.79\pm0.02$ & $2.10\pm0.02$ & $1.94\pm0.02$ & $1.69\pm0.01$ \\\hline
		$n_j \geq 3$ & $2.47\pm0.03$ & $2.33\pm0.03$ & $2.21\pm0.02$ & $2.10\pm0.02$ & $2.02\pm0.02$ & $1.91\pm0.02$ & $1.89\pm0.02$ & $0.92\pm0.01$ & $1.77\pm0.02$ & $1.63\pm0.01$ & $1.42\pm0.01$ \\\hline
		$n_b = 1$ & $1.32\pm0.02$ & $1.20\pm0.02$ & $1.16\pm0.02$ & $1.11\pm0.02$ & $1.04\pm0.01$ & $1.00\pm0.01$ & $0.99\pm0.01$ & $0.92\pm0.01$ & $0.93\pm0.01$ & $0.86\pm0.01$ & $0.736\pm0.009$ \\\hline
		$|m_{t}^{rec} - m_t| < 50$ GeV & $0.74\pm0.02$ & $0.68\pm0.01$ & $0.65\pm0.01$ & $0.63\pm0.01$ & $0.58\pm0.01$ & $0.56\pm0.01$ & $0.54\pm0.01$ & $0.509\pm0.009$ & $0.504\pm0.009$ & $0.454\pm0.008$ & $0.393\pm0.006$ \\\hline
		$n_{\mu^+ \mu^-} = 2$ & $0.74\pm0.02$ & $0.68\pm0.01$ & $0.65\pm0.01$ & $0.63\pm0.01$ & $0.58\pm0.01$ & $0.56\pm0.01$ & $0.54\pm0.01$ & $0.509\pm0.009$ & $0.504\pm0.009$ & $0.454\pm0.008$ & $0.393\pm0.006$ \\\hline
		$m_{total} < 1$ TeV & $0.43\pm0.01$ & $0.39\pm0.01$ & $0.38\pm0.01$ & $0.349\pm0.009$ & $0.306\pm0.008$ & $0.281\pm0.007$ & $0.262\pm0.007$ & $0.248\pm0.006$ & $0.238\pm0.006$ & $0.197\pm0.005$ & $0.151\pm0.004$ \\\hline
		$|m_{S_{1,2}}^{rec} - m_S| < 30$ GeV & $0.42\pm0.01$ & $0.39\pm0.01$ & $0.38\pm0.01$ & $0.347\pm0.009$ & $0.301\pm0.008$ & $0.279\pm0.007$ & $0.260\pm0.007$ & $0.245\pm0.006$ & $0.235\pm0.006$ & $0.193\pm0.005$ & $0.146\pm0.004$ \\\hline
	\end{tabular}
	}
	\caption{Event yields after each cut for the eleven 
benchmark signal points, in the analysis for  $t\to SSu, S\to\mu^+\mu^-$. \mr{We fix $\tilde{Y}_{13} = \tilde{Y}_{31} = 0.1$, $\Lambda = 1$ TeV and $\mathcal{B}(S\to \mu^+\mu^-) = 1$.} The 
event yields presented assume a collected luminosity of $\mathcal{L} = 150~{\rm 
fb}^{-1}$.}
	\label{tab:Third_Signal_Up}
\end{table}

\begin{table}[t]
	\centering
	\scalebox{0.63}{
	\begin{tabular}{|l|c|c|c|c|c|c|c|c|c|c|c|}
		\hline
		Cuts/$m_S$ & 20 GeV & 30 GeV & 40 GeV & 50 GeV & 60 GeV & 70 GeV & 80 GeV & 90 GeV & 100 GeV & 120 GeV & 150 GeV\\\hline
		basic & $0.76\pm0.01$  & $0.90\pm0.01$ & $0.85\pm0.01$ & $0.729\pm0.008$ & $0.609\pm0.006$ & $0.526\pm0.005$ & $0.495\pm0.004$ & $0.308\pm0.003$ & $0.445\pm0.004$ & $0.395\pm0.003$ & $0.192\pm0.002$ \\\hline
		$n_j \geq 3$ & $0.68\pm0.01$  & $0.65\pm0.01$ & $0.624\pm0.009$ & $0.520\pm0.006$ & $0.431\pm0.005$ & $0.369\pm0.004$ & $0.350\pm0.003$ & $0.161\pm0.002$ & $0.315\pm0.003$ & $0.281\pm0.003$ & $0.163\pm0.001$ \\\hline
		$n_b = 1$ & $0.347\pm0.009$ & $0.356\pm0.008$ & $0.325\pm0.006$ & $0.274\pm0.005$ & $0.227\pm0.003$ & $0.194\pm0.003$ & $0.183\pm0.003$ & $0.161\pm0.002$ & $0.165\pm0.002$ & $0.145\pm0.002$ & $0.083\pm0.001$ \\\hline
		$|m_{t}^{rec} - m_t| <$ 50 GeV & $0.219\pm0.007$ & $0.151\pm0.005$ & $0.145\pm0.004$ & $0.127\pm0.003$ & $0.105\pm0.002$ & $0.090\pm0.002$ & $0.086\pm0.002$ & $0.046\pm0.001$ & $0.075\pm0.001$ & $0.065\pm0.001$ & $0.046\pm0.001$ \\\hline
		$n_{\mu^+\mu^-} =2$ & $0.219\pm0.007$ & $0.151\pm0.005$ & $0.145\pm0.004$ & $0.127\pm0.003$ & $0.105\pm0.002$ & $0.090\pm0.002$ & $0.086\pm0.002$ & $0.046\pm0.001$ & $0.075\pm0.001$ & $0.065\pm0.001$ & $0.046\pm0.001$ \\\hline
		$m_{total} < 1$ TeV & $0.186\pm0.007$ & $0.125\pm0.005$ & $0.123\pm0.004$ & $0.101\pm0.003$ & $0.078\pm0.002$ & $0.062\pm0.002$ & $0.057\pm0.001$ & $0.044\pm0.001$ & $0.048\pm0.001$ & $0.039\pm0.001$ & $0.023\pm0.001$ \\\hline
		$|m_{S_{1,2}}^{rec} - m_S| < 30$ GeV  & $0.185\pm0.007$ & $0.121\pm0.005$ & $0.114\pm0.004$ & $0.095\pm0.003$ & $0.074\pm0.002$ & $0.059\pm0.002$ & $0.054\pm0.001$ & $0.044\pm0.001$ & $0.046\pm0.001$ & $0.037\pm0.001$ & $0.022\pm0.001$ \\\hline
	\end{tabular}
	}
	\caption{Event yields after each cut for the eleven
benchmark signal points, in the analysis for  $t\to SSc, S\to\mu^+\mu^-$. \mr{We fix $\tilde{Y}_{23} = \tilde{Y}_{32} = 0.1$, $\Lambda = 1$ TeV and $\mathcal{B}(S\to \mu^+\mu^-) = 1$.} The 
event yields presented assume a collected luminosity of $\mathcal{L} = 150~{\rm 
fb}^{-1}$.}
	\label{tab:Third_Signal_Charm}
\end{table}

\begin{table}[t]
	\centering
	\scalebox{0.75}{
	\begin{tabular}{|l|c|c|c|c|c|c|c|}
		\hline
		Cuts/Background & $tW$ & $t\bar{t}W/t\bar{t}Z$ & $ZZZ/WWZ$ & $ZZ/WZ/WW$ & $t\bar{t}$ ($\mu$) & $tZ$ \\\hline
		basic & $<~42$ & $8.8\pm0.2$ & $1.372\pm0.006$ & $1780\pm727$ & $17\pm5$ &  $0.07\pm0.02$ \\\hline
		$n_j \geq 3$ & --- & $5.6\pm0.2$ & $0.100\pm0.002$ & $297\pm297$ & $7\pm3$ & $0.03\pm0.01$ \\\hline
		$n_b = 1$ & --- & $2.6\pm0.1$ & $0.010\pm0.001$ & $<~4$ & $3\pm2$ &  $0.011\pm0.008$ \\\hline
		$|m_{t}^{rec} - m_t| <$ 50 GeV & --- & $0.62\pm0.06$ & $0.002\pm0.001$ & --- & $<~2$ &  $<~0.006$ \\\hline
		$n_{\mu^+ \mu^-} = 2$ & --- & $0.62\pm0.06$ & $0.002\pm0.001$ & --- & --- &  --- \\\hline
		$m_{total} < 1$ TeV & --- & $0.60\pm0.06$ & $<~0.0003$ & --- & --- & --- \\\hline
	\end{tabular}
	}
	\caption{Event yields after each cut for the dominant backgrounds, in 
the analysis for $t\to SSq, S\to\mu^+\mu^-$. 
The $Z$ + jets sample is reduced to negligible values after the second cut.
The event yields presented assume a collected luminosity of $\mathcal{L} = 
150~{\rm fb}^{-1}$. }
	\label{tab:Third_Background}
\end{table}
\end{landscape}
\begin{figure}[h]
 \centering
  \includegraphics[width=0.49\textwidth]{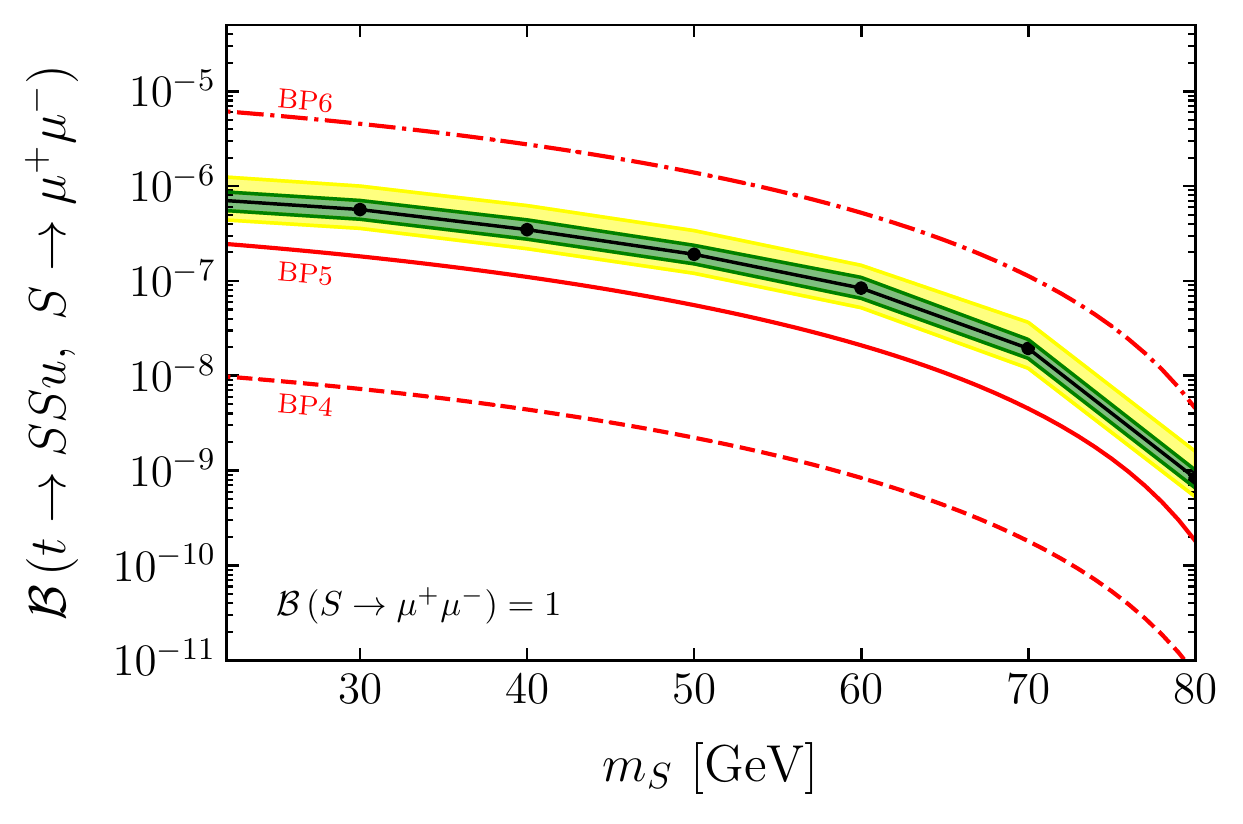}
  \includegraphics[width=0.49\textwidth]{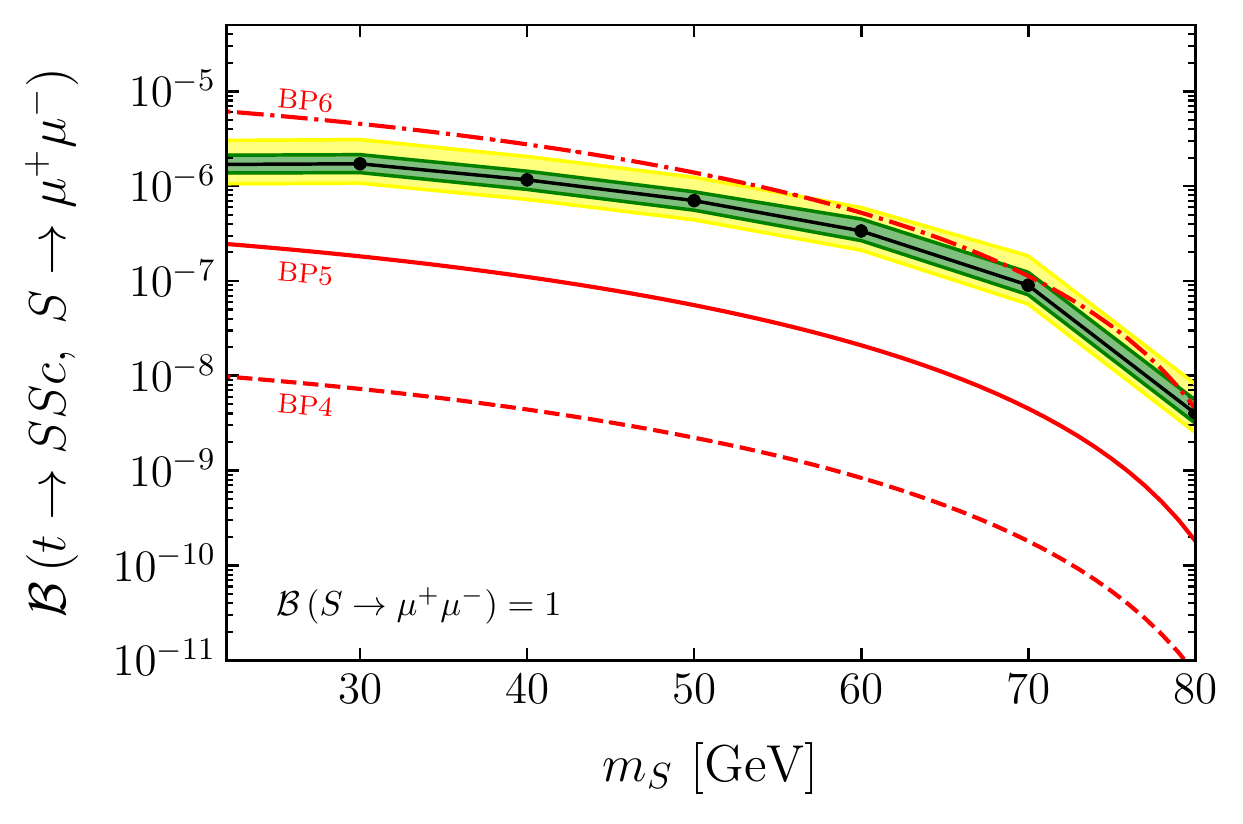}\\
      \includegraphics[width=0.49\textwidth]{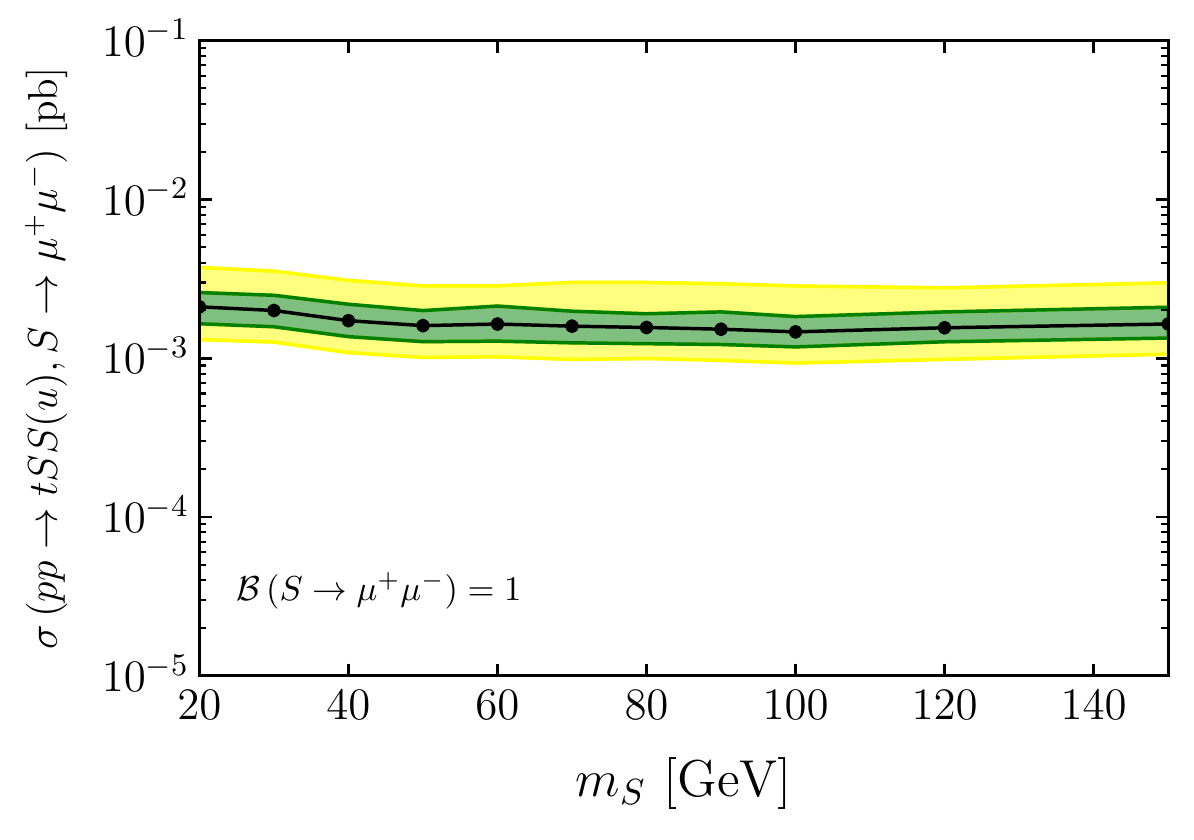}
  \includegraphics[width=0.49\textwidth]{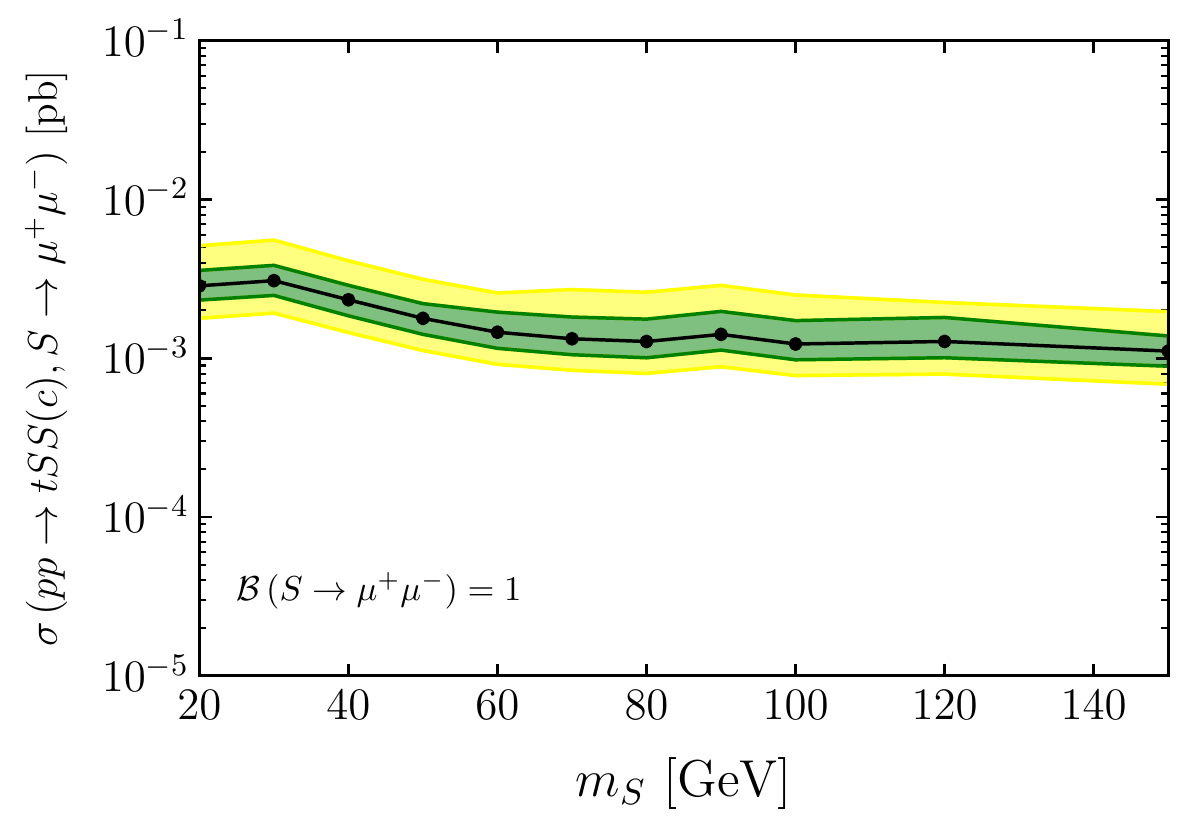}\\
 \caption{In the upper (bottom) panels, we show the 95\% CL limits on the 
branching ratio (cross section times branching ratio) that can be tested in the 
$\mu^+\mu^-$ channel, in the 
analysis proposed for $t\rightarrow S S q,~S\rightarrow \mu^+\mu^-$, with $q = 
u~(c)$ in the panels on the left (right). The green and yellow bands 
show the $\pm 1 \sigma$ and $\pm 2\sigma$ uncertainty on the limits, 
respectively. The limits are obtained for a 
collected luminosity $\mathcal{L} = 150~{\rm fb}^{-1}$. Superimposed are the 
theoretical expectations in three BPs.}
 \label{fig:BR_3mumu}
\end{figure}

\section{Conclusions}
\label{sec:conclusions}
Using an effective field theory approach, we have shown that, in models of new 
physics with light pseudo-scalar singlet degrees of freedom $S$, new top 
flavor-changing neutral currents (FCNCs) arise.  These can trigger the rare top 
decay $t\to Sq$, as well as 
produce the top quark in association with $S$ in proton-proton collisions. At 
the LHC, both processes can be captured by inclusive searches for $pp\to tS + 
j$.

We 
have shown that, if there is a $\mathbb{Z}_2$ symmetry under which $S$ is odd 
while all 
the SM quarks are even, then one expects rather the FCNC process $pp\to 
tSS + j$. Under this hypothesis, $S$ decays exclusively into leptons, 
predominantly into taus and muons.
For completeness we have shown how these ideas can be implemented in a 
concrete composite Higgs model. We have also demonstrated that experimental 
analyses currently performed at the LHC are not significantly sensitive to 
these interactions.
Thus, we have worked out three new dedicated searches in full detail, including 
detector simulation and with a rigorous quantification of uncertainties, to 
probe the processes $pp\to tS, S\to \mu^+ \mu^-$; $pp\to tS, S\to\tau^+ 
\tau^-$ and $pp\to tSS, S\to\mu^+ \mu^-$.

In the channel $pp\to tS, S\to \mu^+ \mu^-$, we focus on final state events with 
three light leptons and jets (with exactly one $b$-tagged). The main 
discriminating variable is the invariant mass of the hardest opposite sign muon 
pair. The dominant 
background ensues from $t\overline{t}$ and $tZ$.
For an integrated luminosity of 150 fb$^{-1}$, we find that  a production cross section $\sigma(pp\to tS, S\to \mu^+ \mu^-) > 10^{-3}$ pb can be tested at the 95\% CL. The highest sensitivity is attained for $m_S \sim$ 150 GeV, for which
\mpr{the maximum number of events compatible with background fluctuations is found to be $\sim 170~(190)$ in the up (charm) channel. Considering a new physics coupling and scale $\Lambda$ equal to $0.1$ and $5$ TeV, we predict $\sim 410~(140)$ signal events, assuming that $\mathcal{B}(S\to \mu^+\mu^-) = 1$. Therefore, this benchmark point could be excluded in the analysis with the up quark. For the same singlet mass, we can probe $\mathcal{B}(t\to S q) > 5 ~ (15) \times 10^{-7}$ at the 95\% CL. In turn, for $\mathcal{O}(1)$ couplings in the UV, these results translate into a 
lower bound on $\Lambda \sim 90$ TeV.}

In the channel $pp\to tS, S\to \tau^+\tau^-$, we focus on final state events with 
one light lepton and jets (with one $b$-tagged and two hadronic taus). The main 
discriminating variable is the invariant mass of the two tau-jets. The dominant 
backgrounds are $t\overline{t}$ and $tW$. 
For an integrated luminosity of 150 fb$^{-1}$, we find that  a production cross section $\sigma(pp\to tS, S\to \tau^+ \tau^-) > 10^{-2}$ pb can be tested at the 95\% CL. The highest sensitivity is attained for $m_S \sim$ 50 GeV, for which \mpr{the maximum number of events is found to be $\sim 2.6~(2.2)\times 10^{3}$ in the up (charm) channel. Considering a new physics coupling and scale equal to $0.1$ and $1$ TeV, we predict $\sim 7.8~(6.5)\times 10^4$ signal events, assuming that $\mathcal{B}(S\to \tau^+\tau^-) = 1$.  Therefore, this benchmark point could be excluded. For the same singlet mass, we can probe $\mathcal{B}(t\to S q) > 11 ~ (12) \times 10^{-6}$ at the 95\% CL. In turn, for $\mathcal{O}(1)$ couplings in the UV, these results translate into a 
lower bound on $\Lambda \sim 75$ TeV.}

For $pp\to tS S, S\to\mu^+\mu^-$, we concentrate on events with four 
light leptons and jets (one of which $b$-tagged). The principal discriminating 
variable is the invariant mass of the four leptons. This search is 
in good approximation background free. 
For an integrated luminosity of 150~fb$^{-1}$, we find that  a production cross section $\sigma(pp\to tS S, S\to \mu^+ \mu^-) > 10^{-3}$ pb can be tested at 95\% CL.
\mpr{For $m_S \sim 80$ GeV, the maximum number of events is found to be $\sim 230~(190)$ in the up (charm) channel. Considering a new physics coupling and scale equal to $1.0$ and $1$ TeV, we predict $\sim 1280~(220)$ signal events, assuming that $\mathcal{B}(S\to \mu^+\mu^-) = 1$. Therefore, this benchmark point could be excluded. The strongest limits in branching ratio, $\mathcal{B}(t\to S S q) > 5~(25) \times 10^{-10}$, are obtained for this mass point, while we can probe $\mathcal{B}(t\to S S q) > 10^{-6}$ at the 95\% CL in the small mass regime.} In turn, for $\mathcal{O}(1)$ couplings in the UV, these results translate into a lower bound on $\Lambda \sim 2$ TeV.

Naive prospects for higher luminosities can be obtained by scaling the statistical 
significance with $\sqrt{\mathcal{L}}$. Thus, at $\mathcal{L}=3$~ab$^{-1}$, we 
expect to probe 
scales of order $200$, $160$, and $3$~TeV in each of the channels, 
respectively.

\section*{Acknowledgements}
The authors were partially funded by FCT - Funda\c c\~ao para a Ci\^encia e a Tecnologia, I.P., Portugal, under projects CERN/FIS-PAR/0024/2019 (MC, NC and MR) and CERN/FIS-PAR/0002/2019 (NC and AP). MC is also supported by the Spanish MINECO under the Juan de la Cierva programme as well as by the Ministry of Science and Innovation under grant numbers FPA2016-78220-C3-3-P (fondos FEDER),
and by the Junta de Andaluc{\'\i}a grants FQM 101 and A-FQM-211-UGR-18
(fondos FEDER). AP acknowledges the 
support by FCT through grant SFRH/BD/129321/2017.
MR acknowledges support by FCT under the grant PD/BD/142773/2018 and by LIP (FCT,
COMPETE2020-Portugal2020, FEDER, POCI-01-0145-FEDER-007334). The computational part of this work was supported by
INCD (funded by FCT and FEDER under the project 01/SAICT/2016 nr.
022153) and by the Minho Advanced Computing Center (MACC).
\noindent

\bibliographystyle{style}
\bibliography{references}

\end{document}